\documentclass[aps,prd,superscriptaddress,onecolumn,floatfix,nofootinbib]{revtex4}
\usepackage{amsmath,amssymb,graphicx,tabularx,url,multirow,hyperref}

\begin{document}

\def\contentsname{{\normalsize Content}}
\def\tablename{Table}
\def\figurename{Figure}

\def\pveto{P_\text{veto}}
\def\nj{n_\text{jets}}
\def\meff{m_\text{eff}}
\def\ptmin{p_T^\text{min}}
\def\gtot{\Gamma_\text{tot}}
\def\as{\alpha_s}
\def\az{\alpha_0}
\def\gz{g_0}
\def\w{\vec{w}}
\def\sdag{\Sigma^{\dag}}
\def\s{\Sigma}
\newcommand{\psib}{\overline{\psi}}
\newcommand{\Psib}{\overline{\Psi}}
\newcommand\one{\leavevmode\hbox{\small1\normalsize\kern-.33em1}}
\newcommand{\Mpl}{M_\mathrm{Pl}}
\newcommand{\p}{\partial}
\newcommand{\mat}{\mathcal{M}}
\newcommand{\lag}{\mathcal{L}}
\newcommand{\ord}{\mathcal{O}}
\newcommand{\ope}{\mathcal{O}}
\newcommand{\qqquad}{\qquad \qquad}
\newcommand{\qqqquad}{\qquad \qquad \qquad}

\newcommand{\qb}{\bar{q}}
\newcommand{\matx}{|\mathcal{M}|^2}
\newcommand{\really}{\stackrel{!}{=}}
\newcommand{\msbar}{\overline{\text{MS}}}
\newcommand{\qns}{f_q^\text{NS}}
\newcommand{\lqcd}{\Lambda_\text{QCD}}
\newcommand{\met}{\slashchar{p}_T}
\newcommand{\pmiss}{\slashchar{\vec{p}}_T}

% for SUSY fit projects
\newcommand{\Mzero}{\ensuremath{m_0}}
\newcommand{\MOneHalf}{\ensuremath{m_{1/2}}}
\newcommand{\omegacdm}{\ensuremath{\Omega h^2}}
\newcommand{\Mtop}{\ensuremath{m_t}}
\newcommand{\Mone}{\ensuremath{M_1}}
\newcommand{\Mtwo}{\ensuremath{M_2}}
\newcommand{\Mthree}{\ensuremath{M_3}}
\newcommand{\Deltaamu}{\ensuremath{\mathrm{\Delta a_\mu}}}
\newcommand{\chisq}{{\ensuremath -2\log L}}
\newcommand{\Alambda}{{\ensuremath {A_\lambda}}}
\newcommand{\Akappa}{{\ensuremath {A_\kappa}}}
\newcommand{\tkappa}{{\ensuremath{\tilde \kappa}}}
\newcommand{\abs}[1]{\left |#1\right |}
\newcommand{\unit}[1]{\ensuremath{\, \mathrm{#1}}}

\newcommand{\lsp}{\tilde \chi}
\newcommand{\sq}{\tilde{q}}
\newcommand{\go}{\tilde{g}}
\newcommand{\st}[1]{\tilde{t}_{#1}}
\newcommand{\stb}[1]{\tilde{t}_{#1}^*}
\newcommand{\nz}[1]{\tilde{\chi}_{#1}^0}
\newcommand{\cp}[1]{\tilde{\chi}_{#1}^+}
\newcommand{\cm}[1]{\tilde{\chi}_{#1}^-}
\newcommand{\CP}{CP}

% all the masses 
\providecommand{\mg}{m_{\tilde{g}}}
\providecommand{\mst}[1]{m_{\tilde{t}_{#1}}}
\newcommand{\msn}[1]{m_{\tilde{\nu}_{#1}}}
\newcommand{\mch}[1]{m_{\tilde{\chi}^+_{#1}}}
\newcommand{\mne}[1]{m_{\tilde{\chi}^0_{#1}}}
\newcommand{\msb}[1]{m_{\tilde{b}_{#1}}}
\newcommand{\vsm}{\ensuremath{v_\text{SM}}}

% units of measure
\newcommand{\cmu}{\text{cm}}
\newcommand{\mev}{\text{MeV}}
\newcommand{\gev}{\text{GeV}}
\newcommand{\tev}{\text{TeV}}
\newcommand{\fb}{\text{fb}}
\newcommand{\ab}{\text{ab}}
\newcommand{\pb}{\text{pb}}
\newcommand{\br}{\text{BR}}
\newcommand{\sign}{\text{sign}}
\newcommand{\iab}{\text{ab}^{-1}}
\newcommand{\ifb}{\text{fb}^{-1}}
\newcommand{\ipb}{\text{pb}^{-1}}

% really great macro by Chris Lester
\def\slashchar#1{\setbox0=\hbox{$#1$}           % set a box for #1
   \dimen0=\wd0                                 % and get its size
   \setbox1=\hbox{/} \dimen1=\wd1               % get size of /
   \ifdim\dimen0>\dimen1                        % #1 is bigger
      \rlap{\hbox to \dimen0{\hfil/\hfil}}      % so center / in box
      #1                                        % and print #1
   \else                                        % / is bigger
      \rlap{\hbox to \dimen1{\hfil$#1$\hfil}}   % so center #1
      /                                         % and print /
   \fi}
\newcommand{\dslash}{\slashchar{\partial}}
\newcommand{\Dslash}{\slashchar{D}}

\newcommand{\eg}{\textsl{e.g.}\;}
\newcommand{\ie}{\textsl{i.e.}\;}
\newcommand{\etal}{\textsl{et al}\;}
%\DeclareMathOperator{\tr}{Tr}

% maximal number of floating environments on each page 
\setlength{\floatsep}{0pt}
\setcounter{topnumber}{1}
\setcounter{bottomnumber}{1}
\setcounter{totalnumber}{1}
\renewcommand{\topfraction}{1.0}
\renewcommand{\bottomfraction}{1.0}
\renewcommand{\textfraction}{0.0}
\renewcommand{\thefootnote}{\fnsymbol{footnote}}

\newcommand{\rig}{\rightarrow}
\newcommand{\lrig}{\longrightarrow}
\renewcommand{\d}{{\mathrm{d}}}
\newcommand{\be}{\begin{eqnarray*}}
\newcommand{\ee}{\end{eqnarray*}}
\newcommand{\gl}[1]{(\ref{#1})}
\newcommand{\ta}[2]{ \frac{ {\mathrm{d}} #1 } {{\mathrm{d}} #2}}
\newcommand{\bee}{\begin{eqnarray}}
\newcommand{\eee}{\end{eqnarray}}
\newcommand{\beeq}{\begin{equation}}
\newcommand{\eeeq}{\end{equation}}
\newcommand{\mc}{\mathcal}
\newcommand{\mr}{\mathrm}
\newcommand{\ep}{\varepsilon}
\newcommand{\eps}{\epsilon}
\newcommand{\emt}{$\times 10^{-3}$}
\newcommand{\emfo}{$\times 10^{-4}$}
\newcommand{\emfi}{$\times 10^{-5}$}

\newcommand{\revision}[1]{{\bf{}#1}}

\newcommand{\hzero}{\ensuremath{h^0}}
\newcommand{\Hzero}{\ensuremath{H^0}}
\newcommand{\Azero}{\ensuremath{A^0}}
\newcommand{\PHiggs}{\ensuremath{H}}
\newcommand{\PW}{\ensuremath{W}}
\newcommand{\PZ}{\ensuremath{Z}}

\newcommand{\sw}{\ensuremath{s_w}}
\newcommand{\cw}{\ensuremath{c_w}}
\newcommand{\swd}{\ensuremath{s^2_w}}
\newcommand{\cwd}{\ensuremath{c^2_w}}

%% 2HDM Higgs masses
\newcommand{\mhhd}{\ensuremath{m^2_{\Hzero}}}
\newcommand{\mhh}{\ensuremath{m_{\Hzero}}}
\newcommand{\mlhd}{\ensuremath{m^2_{\hzero}}}
\newcommand{\Mlh}{\ensuremath{m_{\hzero}}}
\newcommand{\mad}{\ensuremath{m^2_{\Azero}}}
\newcommand{\mhpd}{\ensuremath{m^2_{\PHiggs^{\pm}}}}
\newcommand{\mhp}{\ensuremath{m_{\PHiggs^{\pm}}}}

\newcommand{\sa}{\ensuremath{\sin\alpha}}
\newcommand{\ca}{\ensuremath{\cos\alpha}}
\newcommand{\cad}{\ensuremath{\cos^2\alpha}}
\newcommand{\sad}{\ensuremath{\sin^2\alpha}}
\newcommand{\sbd}{\ensuremath{\sin^2\beta}}
\newcommand{\cbd}{\ensuremath{\cos^2\beta}}
\newcommand{\cb}{\ensuremath{\cos\beta}}
\renewcommand{\sb}{\ensuremath{\sin\beta}}
\newcommand{\tanbd}{\ensuremath{\tan^2\beta}}
\newcommand{\cotbd}{\ensuremath{\cot^2\beta}}
\newcommand{\tanb}{\ensuremath{\tan\beta}}
\newcommand{\tb}{\ensuremath{\tan\beta}}
\newcommand{\cotb}{\ensuremath{\cot\beta}}

\title{Invisible Higgs Decays to Hooperons in the NMSSM}

\author{Anja Butter}
\affiliation{Institut f\"ur Theoretische Physik, Universit\"at Heidelberg, Germany}
\affiliation{LAL, CNRS/IN2P3, Orsay Cedex, France}

\author{Tilman Plehn}
\affiliation{Institut f\"ur Theoretische Physik, Universit\"at Heidelberg, Germany}

\author{Michael Rauch}
\affiliation{Institute for Theoretical Physics, Karlsruhe Institute of Technology (KIT), Karlsruhe, Germany}

\author{Dirk Zerwas}
\affiliation{LAL, CNRS/IN2P3, Orsay Cedex, France}

\author{Sophie Henrot-Versill\'e}
\affiliation{LAL, CNRS/IN2P3, Orsay Cedex, France}

\author{R\'emi Lafaye}
\affiliation{LAPP, Universit\'e de Savoie, IN2P3/CNRS, Annecy, France}

\begin{abstract}
  The galactic center excess of gamma ray photons can be naturally
  explained by light Majorana fermions in combination with a
  pseudoscalar mediator. The NMSSM provides exactly these
  ingredients. We show that for neutralinos with a significant
  singlino component the galactic center excess can be linked to
  invisible decays of the Standard-Model-like Higgs at the LHC. We
  find predictions for invisible Higgs branching ratios in excess of 50
  percent, easily accessible at the LHC. Constraining the NMSSM
  through GUT-scale boundary conditions only slightly affects this
  expectation. Our results complement earlier NMSSM studies of the
  galactic center excess, which link it to heavy Higgs searches at the
  LHC.
\end{abstract}

\maketitle
\tableofcontents
\newpage

%%%%%%%%%%%%%%%%%%%%%%%%%%%%%%%%%%%%%%%%%%%%%%%%%%%%%%%%%%%%%%%%%%%%%%%%%%%%%%%%
\section{Introduction and basics}
\label{sec:intro}

While the existence of cold dark matter as the main matter component
of today's Universe is generally acknowledged, the particle nature of
it is still elusive. Searches for dark matter coupled to Standard
Model fields with more than a gravitational interaction strength
follow three distinct strategies: direct detection, indirect
detection, and production at colliders. The latter will receive a
significant boost with the start of LHC Run~II. The key question is
how in the case of weakly interacting dark matter the different search
strategies can support and inspire each other. One of the main search
strategies for dark matter at the LHC are invisible Higgs decays, most
notably in weak boson fusion~\cite{eboli_zeppenfeld,jamie}. For
example, in models without new strongly interacting particles such
invisible Higgs decays will drive mono-jet searches and are likely to
dominate over dark matter pair production in weak boson
fusion~\cite{wbf_susy}. In this paper we will reinforce the link
between the specific Fermi galactic center
excess~\cite{hooperon,hooperon_planck, lat_dama} and invisible Higgs decays at
the LHC~\cite{hinv_atlas,hinv_cms} in the framework of the
next-to-minimal supersymmetric Standard Model
(NMSSM)~\cite{nmssm_orig,nmssm_review}.\bigskip

The Fermi gamma ray space telescope searches for dark matter signals
in its photon spectrum. An excess of gamma rays from the galactic
center has for many years avoided possible background
interpretations. It can be explained by annihilating dark matter with
a spherical distribution around the center of our galaxy. Its spectrum
gives preferred mass values for different dark matter candidates. For
an annihilation to bottom quarks the preferred mass of the dark matter
agent (Hooperon) ranges around 40~GeV~\cite{hooperon,hooperon2},
extending all the way to 70~GeV~\cite{calore_cholis_weniger}. In our
analysis of the $b\bar{b}$ case we will follow
Ref.~\cite{calore_cholis_weniger} and assume a conservative LSP of
30~GeV to 70~GeV. The cross section should be in the range of $\sigma
v \approx 1.8 \cdot 10^{-26} \cmu^3/\text{s}$~\cite{hooperon,calore_cholis_weniger}, 
with appropriate theoretical or parametric uncertainties for example from
the choice of dark matter profile, consistent with the latest Planck thermally averaged results~\cite{hooperon_planck}. Such values are intriguingly close to
the expectations for a thermally produced weakly interacting dark
matter particle (WIMP)~\cite{wimp,nimatron}.\bigskip

In the MSSM the preferred mass range of the Hooperon is a challenge
and typically relies on dark matter annihilation into a pair of gauge
boson~\cite{hooperon_mssm}. In the absence of the highly efficient
annihilation through an $s$-channel mediator decaying for example into
$b\bar{b}$ pairs the predicted relic density in the Universe tends to
be too large. Finding efficient annihilation channels is a serious
issue in supersymmetric models~\cite{sfitter_planck,jay,DMpapers}:
first, $s$-channel annihilation through the $Z$-pole, the SM-like
Higgs resonance $H_{125}$, or a heavy Higgs resonance are either
forbidden by other constraints or too small. Any co-annihilation
channel requires an additional supersymmetric particle within 10\% of
the LSP
mass~\cite{stau-co-annihilation,char-co-annihilation,stop-co-annihilation},
which is disfavored by LEP constraints~\cite{lep_constraints}. The
way out is an additional mediator, ideally a pseudoscalar with a mass
not far above twice the LSP mass~\cite{hooperon_extendedmssm}. This
feature is clearly visible in an analysis in terms of simplified
models or effective field theory~\cite{hooperon_simplified}.\bigskip

As an extension of the MSSM, the NMSSM provides exactly such a
mediator, the pseudoscalar part of the singlet/singlino superfield
mixed with the MSSM-like pseudoscalar~\cite{nmssm_singlino}. In the
required mass range it will naturally decay to $b\bar{b}$ pairs, and
with a reduced branching ratio to $\tau^+ \tau^-$. Such an NMSSM setup
can be tested in a parameter scan~\cite{hooperon_scan} and then linked
for example to 4-body Higgs decay~\cite{hooperon_felix}, trilepton
searches at the LHC~\cite{hooperon_china} or even the electroweak
phase transition~\cite{hooperon_phasetransition}.  Because of the
structure of the NMSSM we can follow two strategies to accommodate the
galactic center excess~\cite{papucci_zurek,hooperon_bino}: first, we
can keep the standard bino--Higgsino LSP composition of the MSSM and
only couple the neutralinos to the light additional
pseudoscalar. Alternatively, we can replace the bino content by a
singlino content and assume a singlino--Higgsino, or better
bino--singlino--Higgsino LSP. Again, it will couple to the
pseudoscalar mediator.

For the bino--Higgsino case the relevant LSP and mediator states are
not decoupled from the Standard Model. This means that for example the
pseudoscalar mediator can be searched for at the
LHC~\cite{papucci_zurek}. The singlino--Higgsino channel is more
challenging. After introducing the NMSSM singlet/singlino sector and
its phenomenology including some useful formulas, in
Sec.~\ref{sec:hooperon} we will study its TeV-scale parameter space
linked to the galactic center excess. In Sec.~\ref{sec:hinv} we will
link the Hooperon parameter space to the size of invisible Higgs
couplings. It will turn out that similar to a dark matter Higgs
portal~\cite{portal} the NMSSM interpretation of the galactic center
excess will lead to invisible Higgs decays with branching ratios
accessible during the upcoming LHC run. In Sec.~\ref{sec:high_scale}
we will apply the same criteria to a high-scale NMSSM setup. For this
model a global \textsc{SFitter} likelihood analysis is useful, before
we turn to the link between the galactic center excess and invisible
Higgs decays.

%%%%%%%%%%%%%%%%%%%%%%%%%%%%%%%%%%%%%%%%%%%%%%%%%%%%%%%%%%%%%%%%%%%%%%%%%%%%%%%%
\subsection{NMSSM}
\label{sec:nmssm}

Compared to the minimal supersymmetric Standard Model the
superpotential of the NMSSM~\cite{nmssm_orig,nmssm_review} includes an
additional singlet superfield $\hat{S}$ and the associated terms
\begin{align}
W_\text{NMSSM} = W_\text{MSSM} 
 + \lambda \, \hat{S} \hat{H}_u \hat{H}_d 
 + \xi_F \hat{S} 
 + \dfrac{\mu'}{2} \, \hat{S}^2 
 + \dfrac{\kappa}{3} \, \hat{S}^3 \; ,
\label{eq:WNMSSM}
\end{align}
where $\lambda$ and $\kappa$ are dimensionless couplings coupling the
singlet to itself and to the Higgs bosons. When the singlet acquires a
vacuum expectation value $v_s$, the Higgs--singlet mixing introduces
an effective $\mu$-term $\mu_\text{eff}=\lambda v_s$. The quadratic
term proportional to $\mu'$ is the supersymmetric mass term for the
singlet, comparable to the $\mu$-term for the MSSM Higgs
bosons. Assuming a global supersymmetry the tadpole term proportional
to $\xi_F$ can be removed though a constant shift of the singlet
field.  Finally, with the help of an ad-hoc $\mathbb{Z}_3$-symmetry we
can make the superpotential scale invariant and set the one remaining
dimensionful parameter, $\mu'$, to zero.

The extended superpotential in Eq.\eqref{eq:WNMSSM} in terms of the
superfield $\hat{S}$ can be translated into additional
soft-SUSY-breaking terms for the physical singlet field
$S$~\cite{nmssm_review},
\begin{align}
-\mathcal{L}_\text{soft}^\text{NMSSM} 
=  m_S^2  \abs{S}^2 
+ \left(  \lambda \Alambda H_u H_d S 
        + \dfrac{\kappa}{3} \, \Akappa S^3 
        + \dfrac{m_S^{'2}}{2} \, S^2 
        + \xi_S S 
        + \text{h.c.} 
  \right) \; .
\end{align} 
The $A_{\lambda, \kappa}$ carry mass dimension and fix the scale of
the Lagrangian, while $\lambda$ and $\kappa$ defined in
Eq.\eqref{eq:WNMSSM} are c-numbers.  An alternative parametrization of
the same Lagrangian uses the mass terms $m_3^2=B\mu$ and
$m_S^{'2}=B'\mu'$.  To be consistent with the $\mathbb{Z}_3$-symmetry
of the superpotential we also eliminate the corresponding
SUSY-breaking terms by setting $m_3^2 = m_S^{'2}=\xi_S=0$.  In the
presence of the effective $\mu$-term we can neglect the original $\mu$
parameter, eliminating yet additional independent scale in the
Lagrangian. Correspondingly, $\mu$ will in the following indicate the
effective $\mu$ term.  The relevant NMSSM Lagrangian now reads
\begin{align}
-\mathcal{L}_\text{soft}^\text{NMSSM} 
=  m_S^2 \, \abs{S}^2 
+ \left(  \lambda \Alambda H_u H_d S 
        + \dfrac{\kappa}{3} \,  \Akappa S^3 
        + \text{h.c.} 
  \right) \; .
\label{eq:nmssm_lag}
\end{align} 
\bigskip

In the MSSM, the minimization conditions of the Higgs potential can be
used to replace $m_{H_u}^2$ and $m_{H_d}^2$ by $m_Z$ and $\tan \beta$
in the broken phase. Using the additional minimization condition of
the NMSSM $m_S^2$ can be expressed in terms of $\mu$. The
Higgs--singlet sector~\cite{nmssm_singlino} is therefore fully described
by the parameters $\lambda, \kappa, \Alambda, \Akappa, \mu, \tan
\beta$, and the mass of the $Z$-boson.

For specific NMSSM models we have to define the input scale of these
parameters. The ratio of the VEVs $\tan \beta$ is always evaluated at
the weak scale $m_Z$, because it assumes electroweak symmetry
breaking. For the high-scale models discussed in
Sec.~\ref{sec:high_scale} $\lambda, \kappa$ and $\mu$ are set at the
SUSY scale of 1~TeV, while $\Alambda$ and $\Akappa$ can either be unified to
$A_0$ at the GUT scale or set individually (also at the GUT
scale). For the low-scale models in Sec.~\ref{sec:low_scale} all
supersymmetric parameters including $\lambda, \kappa, \Alambda,
\Akappa, \mu$, the squark and slepton masses, etc. are set at the SUSY
scale.

%%%%%%%%%%%%%%%%%%%%%%%%%%%%%%%%%%%%%%%%%%%%%%%%%%%%%%%%%%%%%%%%%%%%%%%%%%%%%%%%
\subsection{Higgs--singlet--singlino sector}
\label{sec:higgs_singlet}

Compared to the minimal supersymmetric Higgs sector of the MSSM, the
phenomenology of the NMSSM is strongly modified by the additional
particles, a scalar and a pseudoscalar Higgs bosons and a fifth
neutralino. While in general the mass of the singlet states is a free
parameter, we will assume that the singlino contributes to a light LSP
and that the singlet Higgs states are therefore lighter than their
SM-like counterparts. For example, the SM-like Higgs boson with its
mass of 125~GeV will typically be the second-lightest CP-even Higgs
scalar. For a scale-invariant superpotential we can write out the
symmetric Higgs mass matrix~\cite{NMSSM_higgs_calc,
  NMSSM_higgs_calc_twoloop} in the $(H,h,S)$-basis, where $h$ is the
SM-like Higgs boson
\begin{align}
M_{H,h,S}^2= m_Z^2
\begin{pmatrix}
  s^2_{2\beta} \left( 1- \dfrac{\lambda^2}{g^2} \right) 
+ \dfrac{2\mu}{s_{2\beta}m_Z^2} \left( \Alambda + \tkappa \mu\right) 
&  \qquad c_{2 \beta} s_{2\beta}  \left( 1- \dfrac{\lambda^2}{g^2} \right) 
& - c_{2 \beta} \dfrac{\lambda}{g m_Z} 
                     \left( \Alambda+ \tkappa \mu\right) \\
\cdot 
& c^2_{2 \beta} + s^2_{2\beta} \dfrac{\lambda^2}{g^2}
& \dfrac{2 \lambda}{g m_Z} 
  \left(\mu- s_{2 \beta} \dfrac{\Alambda}{2}
        + s_{2 \beta} \tkappa \mu\right) \\
\cdot & \cdot 
& s_{2\beta} \dfrac{\lambda^2 \Alambda}{2 g^2 \mu} 
  + \dfrac{\tkappa  \mu}{m_Z^2} \left(\Akappa+4\tkappa \mu\right)
\end{pmatrix}
\label{eq:higgs_matrix}
\end{align}
using the usual notation $s_\beta = \sin \beta$ and $c_\beta = \cos
\beta$. In terms of the ordered mass eigenstates $H_2 \equiv 
H_{125}$ means that throughout our analysis we identify the
second-lightest Higgs with the observed SM-like state. Instead of
$\kappa$ and $\lambda$, the modified parameter set 
\begin{align}
\phantom{haalllooooooo}
\tkappa =& \dfrac{\kappa}{\lambda} 
\quad && \text{(singlet mass parameter)} \notag \\
&\dfrac{\lambda}{g} \; 
\quad && \text{(singlet decoupling parameter)}
\phantom{haalllooooooo}
\label{eq:def_tilde}
\end{align}
appears in the diagonal entries for the light and heavy MSSM-like
Higgs states. In the following, we replace $\kappa$ with $\tkappa$ but
keep $\lambda$ instead of trivially rescaling it by a constant $g$. At
tree level the two NMSSM parameters take the pressure off the stop
sector for small values of $\tan \beta$. In our basis conventions the
second Higgs state it the SM-like observed resonance. This means we
can decouple the singlet contributions from the observed Higgs. Setting
\begin{align}
\Alambda 
= 2 \mu \left( \dfrac{1}{s_{2 \beta}} - \tkappa \right)
%= 2 \mu \left( \dfrac{1}{s_{2 \beta}} - \dfrac{\kappa}{\lambda} \right)
\label{eq:decoup_higgs}
\end{align}
removes the (2,3) entry from the mass matrix and therefore decouples
the singlet sector from the SM-like Higgs boson $h^0$. Note that this
condition does not require any of the couplings in the NMSSM
potential of Eq.\eqref{eq:nmssm_lag} to vanish. 

More generally, we can decouple the singlet from all other Higgs
states in the limit $\lambda \ll g < 1$. This way the two
corresponding entries in the extended Higgs mass matrix vanish. To
make the singlet itself heavy we need to increase its entry in the
mass matrix in the limit $\lambda \ll g$. Neglecting $\Akappa$, the
singlet entry in the Higgs mass matrix is $(2 \tkappa \mu)^2$, which
for finite $\kappa$ consistently decouples with the single condition
$\lambda \ll 1$.\bigskip

Aside from the extra CP-even Higgs, the singlet extensions of the MSSM
Higgs sector adds an additional pseudoscalar.  We can transform the
weak eigen-basis $\left(H_u, H_d,S\right)$ into a mass basis
$\left(A,S\right)$ by a rotation, so that $A= c_\beta H_u + s_\beta
H_d$. For large values of $\tan \beta$ the mass eigenstate $A$ is
approximately given by $H_d$. Removing the Goldstone modes the
$3\times3$ mass matrix in terms of the weak eigenstates can be reduced
to a $2 \times 2$ mass matrix in the basis ($A,S$), which reads
\begin{align}
M_{A,S}^2= m_Z^2
\begin{pmatrix}
  \dfrac{2 \mu \left( \Alambda+\tkappa \mu\right)}{s_{2\beta} m_Z^2} 
& \dfrac{\lambda}{g m_Z} \left( \Alambda - 2 \tkappa \mu \right)\\
\cdot 
& s_{2\beta} \dfrac{\lambda^2}{g^2} 
    \left( \dfrac{\Alambda}{2\mu} +2 \tkappa \right)
  -3\tkappa \dfrac{\mu \Akappa}{m_Z^2}\\
\end{pmatrix}
\simeq m_Z^2
\begin{pmatrix}
  \dfrac{4 \mu^2}{s_{2\beta}^2 m_Z^2} 
& 2 \dfrac{\lambda}{g m_Z} \dfrac{\mu}{s_{2\beta}} \\
\cdot 
& \dfrac{\lambda^2}{g^2} -3\tkappa \dfrac{\mu \Akappa}{m_Z^2} \\
\end{pmatrix} \; .
\label{eq:pseudo_matrix}
\end{align}
The pseudoscalar mass eigenstates are denoted as $A_1$ and $A_2$.
In the second form we use the singlet decoupling condition
Eq.\eqref{eq:decoup_higgs} and assume $s_{2\beta} \ll1$. As for the
scalar sector, the limit $\lambda \ll g$ decouples the singlet;
its squared mass is then given by $- 3 \tkappa \mu \Akappa$.  The upper
left entry of the matrix then corresponds to the MSSM pseudoscalar
mass, $m_A^2= 2 \mu \left( \Alambda+\tkappa
\mu\right)/s_{2\beta}$. This way we can choose either this MSSM-like
mass or $\Alambda$ as input parameter. Similarly, we can replace
$\Akappa$ with the lower-right entry in $M_{A,S}^2$ as the input
parameter.\bigskip

Finally, the supersymmetric partner of the singlet field, the singlino appears
in the neutralino mass matrix,
\begin{align}
M_{\lsp}=
\begin{pmatrix}
M_1 & 0 & -m_Z c_\beta s_w & m_Z s_\beta s_w & 0 \\
0 & M_2 & m_Z c_\beta c_w & -m_Z s_\beta c_w & 0 \\ 
-m_Z c_\beta s_w & m_Z c_\beta c_w & 0 & -\mu & -m_Z s_\beta \dfrac{\lambda}{g} \\
m_Z s_\beta s_w & -m_Z s_\beta c_w & -\mu & 0 & -m_Z c_\beta \dfrac{\lambda}{g} \\
0 & 0 & -m_Z s_\beta \dfrac{\lambda}{g} & -m_Z c_\beta \dfrac{\lambda}{g} & 2 \tkappa \mu
\end{pmatrix}
\label{eq:neut_matrix}
\end{align}
The bottom-right entry indicates that in accordance with
Eq.\eqref{eq:def_tilde} the combination $2 \tkappa \mu$ determines the
singlino mass.  The gauginos do not mix with the singlino. To
altogether decouple the singlino, we have to remove the
singlino--Higgsino mixing via $\lambda \ll 1$ and at the same time
make the singlino heavy, $\tkappa \gg 1$. In contrast, for $\tkappa <
1/2$ the LSP will be mostly singlino. In this case the LSP mass
$m_{\lsp}$ is approximately given by the solution to
\begin{align}
2 \tkappa \mu 
&= m_{\lsp} - m_Z^2\dfrac{ \lambda^2}{g^2} \dfrac{ m_{\lsp}
       -\mu s_{2 \beta}}{ m_{\lsp}^2-\mu^2} \notag \\
\Leftrightarrow \qquad
%\dfrac{\kappa}{\lambda} 
%=\dfrac{1}{2 \mu} 
% \left( m_{\lsp} - \lambda^2 v^2 \dfrac{ m_{\lsp}
%       -\mu s_{2 \beta}}{ m_{\lsp}^2-\mu^2}\right)
m_{\lsp}
%&= 2 \tkappa \mu 
%  + \lambda^2 v^2 \dfrac{ m_{\lsp}
%  -\mu s_{2 \beta}}{ m_{\lsp}^2-\mu^2} 
& \simeq 2 \tkappa \mu 
  + \dfrac{\lambda^2}{g^2} \, \dfrac{m_Z^2}{\mu} \, \dfrac{ 2 \tkappa 
  - s_{2 \beta}}{ 4 \tkappa^2 - 1} \; ,
\label{eq:mass_approx}
\end{align}
so that the LSP mass can be fixed by adjusting $\tkappa$.\bigskip

For the interpretation of the galactic center excess a light
pseudoscalar will be crucial. To describe its relevant couplings we
have to rely on the different mixing matrices. The neutralino mass
matrix will be rotated into its mass eigenstates through a matrix
$(N_{ij})$ with $i,j=1...5$. To rotate the pseudoscalar mass matrix
into its mass eigenstates we also have to consider the Goldstone
mode. The corresponding mixing matrix is $(P_{ij})$ with $i=1,2$ and
$j=1,2,3$ because the Goldstone is not counted as part of the mass
eigenstates $A_{1,2}$. The lighter pseudoscalar Yukawa coupling to
bottom quarks is given by
\begin{align}
g_{A_1 b b}
=i \dfrac{m_b}{\sqrt{2} v c_\beta} P_{11} \; ,
\label{eq:abb}
\end{align}
where $P_{11}$ is the $H_d$ component of the lightest mass eigenstate.
The coupling to the down-type squarks and sleptons is enhanced by
$1/\cos \beta$, which is relevant for large $\tan \beta$. The coupling
mediating the light pseudoscalar decay into the lightest neutralinos
is given by
\begin{align}
g_{A_1 \lsp \lsp}
= &\lambda \sqrt{2} 
   \left(P_{11} N_{14} N_{15}
  + P_{12} N_{13} N_{15}
  + P_{13} N_{13} N_{14} \right)
- \lambda \tkappa \sqrt{2} P_{13} N_{15}^2 \notag \\
&- \left(g_1 N_{11} 
  - g_2 N_{12} \right)
  \left( P_{11} N_{13}
  - P_{12} N_{14}\right) \; .
\label{eq:achichi}
\end{align}
When we assign the second Higgs to be SM-like the lightest
pseudoscalar will be mainly singlet. Therefore the coupling simplifies
to
\begin{align}
g_{A_1 \lsp \lsp}
= &\lambda \sqrt{2}  \left(N_{13} N_{14}  - \tkappa  N_{15}^2 \right) \; ,
\label{eq:achichi_simpl}
\end{align}
where we set $P_{11}=P_{12} \ll P_{13} \simeq 1$.  As $N_{13}$
and $N_{14}$ differ in sign, both contributions will add up. For a
singlino LSP we have sizeable $N_{15} \to 1$, but following
Eq.\eqref{eq:mass_approx} $\tkappa$ ranges around
$m_{\lsp}/(2\mu)$. This means that the singlino term in $g_{A_1 \lsp
  \lsp}$ decreases with increasing $\mu$, but the same is true for the
Higgsino fractions $N_{13}$ and $N_{14}$. Altogether, a large
mediator coupling to the LSP points to a singlino LSP.

%%%%%%%%%%%%%%%%%%%%%%%%%%%%%%%%%%%%%%%%%%%%%%%%%%%%%%%%%%%%%%%%%%%%%%%%%%%%%%%%
\subsection{Dark matter annihilation}
\label{sec:annihilation}

The lightest neutralino being the lightest supersymmetric particle of
the NMSSM provides an excellent dark matter candidate. The key
constraint for our parameter study will be its relic density assuming
thermal production, leaving us with distinct choices of annihilation
mechanisms~\cite{jay, sfitter_planck}. First, with a dark matter mass
around 40~GeV typical co-annihilation
channels~\cite{stau-co-annihilation,char-co-annihilation,stop-co-annihilation}
with particles coupling to the $Z$-boson are constrained by $Z$-pole
measurements at LEP. More generally, they are excluded by direct LEP
searches, unless their masses are carefully tuned to only produce soft
particles in their production and decays. What is left is
neutralino--neutralino annihilation through an $s$-channel mediator:
\begin{itemize}
\item $Z$-funnel annihilation through a Higgsino component in the interaction
\begin{align}
g_{Z \lsp \lsp}
=\dfrac{g}{2 \cos \theta_W} \gamma_\mu \gamma_5 
 \left[ N_{13}N_{13}-N_{14}N_{14}\right] \; .
\label{eq:z_coup}
\end{align}
  Because of the velocity dependence of the annihilation rate $\langle
  \sigma v \rangle$ this channel usually prefers LSP masses slightly
  above or below 45~GeV, because directly on the $Z$-pole the
  annihilation is too efficient.
\item scalar $H_{125}$-funnel annihilation, where the mass of the LSP
  has to strictly speaking be around 63~GeV. However, in combination
  with other annihilation channels the $H_{125}$-funnel can give the
  largest contribution already for LSP masses around 55~GeV. The
  coupling to the Higgs can be found in Eq.\eqref{eq:higgs_coup}. The
  $H \lsp \lsp$ coupling will be relevant for invisible Higgs decays,
  which is why we will discuss it in detail in Sec.~\ref{sec:hinv}.
\item pseudoscalar $A$-funnel annihilation. Unlike in the MSSM we now
  have a singlet and a Higgs pseudoscalar channel. Both of them can
  lead to a highly efficient annihilation with $\sigma v \propto
  v^0$. If the pseudoscalar is mainly singlet, the relevant
  contributions to the neutralino coupling in Eq.\eqref{eq:achichi}
  reduces to
\begin{align}
g_{A \lsp \lsp} 
%=  \lambda \sqrt{2} N_{13} N_{14} 
%- \kappa \sqrt{2} N_{15}^2
= \sqrt{2}  \lambda \left( N_{13} N_{14} 
- \tkappa N_{15}^2 \right) \; .
\label{eq:a_coup}
\end{align}
   The Higgsino components differ in sign, so that the absolute value
   of the couplings add up. Large Higgsino and singlino components therefore lead
   to a strong coupling.
\end{itemize}
%

%%%%%%%%%%%%%%%%%%%%%%%%%%%%%%%%%%%%%%%%%%%%%%%%%%%%%%%%%%%%%%%%%%%%%%%%%%%%%%%%
\subsection{Data and tools}
\label{sec:data}

%----------------------------------------------------
\begin{table}[b!]
  \begin{tabular}{l l l}
    \hline
    measurement & value and error \\
    \hline
    $m_{H_{125}}$ & $(125.09\pm 0.21_\text{stat} \pm 0.11_\text{syst}\pm3.0_\text{theo}  )~\gev$ & \cite{hmass,m_h} \\
    $\Omega_{\lsp} h^2$ & $0.1188 \pm 0.0010_\text{stat} \pm 0.0120 _\text{theo}$ & \cite{hooperon_planck}\\
    $a_{\mu}$ & $(287 \pm 63_\text{exp}  \pm 49_\text{SM} \pm 20_\text{theo}   )\cdot 10 ^{-11}$ & \cite{Amu}\\
    $\br(B\rightarrow X_s \gamma)$ & $ (3.43 \pm 0.21 _\text{stat} \pm 0.07 _\text{syst}) \cdot 10^{-4}$ & \cite{XsGamma}\\
    $\br(B_s^0 \rightarrow \mu^+ \mu^-) $ & $ (3.2 \pm 1.4 _\text{stat} \pm 0.5 _\text{syst} \pm 0.2 _\text{theo}  )\cdot 10^{-9}$   & \cite{Bsmumu} \\
    $\br(B^+ \rightarrow \tau^+ \nu) $ & $ (1.41  \pm 0.43 _\text{stat}) \cdot 10^{-4}$ & \cite{Asner:2010qj}\\
    $\Delta m_{B^0}    $ & $ (0.510     \pm 0.004 _\text{stat}  \pm 0.003 _\text{syst} \pm 0.400_\text{theo}  ) \cdot 10^{12}  \unit{\hbar s}^{-1}$ & \cite{Asner:2010qj}\\
    $\Delta m_{B^0_s}    $ & $ (17.69    \pm 0.08  _\text{stat}    \pm   7.00_\text{theo}  ) \cdot 10^{12} \unit{\hbar s}^{-1}$ & \cite{Asner:2010qj}\\
    $\Gamma_{Z \rightarrow \text{inv}}  $ & $ (-1.9 \pm 1.5  _\text{stat}    \pm  0.2_\text{theo})~\mev  $ & \cite{Zwidth}\\
    $\Gamma_{Z \rightarrow \text{Higgs}}  $ & $ (6.5 \pm 2.3 _\text{stat} \pm 1.0_\text{theo})~\mev    $& \cite{Zwidth} \\
    $m_t$ & $(173.5 \pm 0.6 _\text{stat} \pm 0.8 _\text{syst})~\gev$ & \cite{Topmass}\\
		$m_{\chi^+_1}$ & $>103~\gev$ & \cite{lep_constraints}\\
    $\sigma_{\lsp N}$ & direct detection limits & \cite{xenon} \\
    \hline
  \end{tabular}
  \caption{Data used for the fit including their systematic and
    statistical errors from the measurements and theoretical errors
    for SUSY calculations as far as they are considered.}
  \label{tab:data}
\end{table}
%----------------------------------------------------

In addition to the dark matter relic density and the observed Higgs
mass value there exists a wealth of measurements which might constrain
supersymmetric models.  A large part of the data which we use for our
\textsc{SFitter} parameter study are listed in
Tab.~\ref{tab:data}.

The invisible width for $Z$-decays, $\Gamma_{Z \rightarrow
  \text{inv}}$, is identified with the additional contribution to the
$Z$-width from decays into a pair of LSPs. For LSP candidates with a
mass smaller than 45~GeV, the LEP results for the $Z$-width can be
powerful constraints. In addition, we require the lightest chargino to
have a mass above 103~GeV, because it is very hard to avoid the LEP2
constraints for charged particle production~\cite{lep_constraints}.
This constraint becomes important when we consider regions with small
$\mu<200$~GeV. If the sum of the lightest CP-even and CP-odd Higgs is
smaller than the mass of the $Z$-boson, the total width gets an
additional contribution. This contribution is compared to the
difference between the SM prediction and the measured total width of
the $Z$. The mass of the top is an input parameter to the
supersymmetric SM-like Higgs. The measurement of $a_\mu$ can only be
satisfied with small slepton masses around 400~GeV or lower. As we
decouple the sfermion sector this measurement will only lead to an
overall constant contribution to the log-likelihood. Similarly, with
the decoupled stop sector the exact value of the SM-like Higgs mass is
irrelevant to our analysis, because it can always be re-adjusted
independently of our parameters of
interest~\cite{NMSSM_higgs_calc_twoloop}. Finally we also include the
Xenon100~\cite{xenon} limits on direct detection.  They are most
powerful in the 30 to 100~GeV range, so this measurement will prove to
have a strong exclusion power for the scenarios we are interested in.

The Hooperon as an explanation of the galactic center
excess~\cite{hooperon} is described by two parameters: the LSP mass
and the annihilation cross section in the center of the galaxy. For
our analysis we will assume the conservative
range~\cite{calore_cholis_weniger}
\begin{align}
m_{\lsp} =& (30 \cdots 70) \, \gev \notag \\
\sigma v =& (0.4 \cdots 10) \cdot 10^{-26} \, \frac{\cmu^3}{\text{s}}
\label{eq:hooperon_paras}
\end{align}
A more detailed analysis in the two-dimensional plane will be part of
the full analysis in Sec.~\ref{sec:hinv}.\bigskip

The \textsc{SFitter} fit then determines multidimensional likelihood
maps for the model parameter space.  A set of Markov chains selects
points in the model space following a Breit--Wigner proposal
function. For each point we compute all considered observables and
determine a generalized $\chi^2$
value~\cite{sfitter_release,sfitter_susy,fittino}. Theoretical and
experimental errors are combined using the \textsc{RFit}
scheme~\cite{rfit}, which corresponds to a consistent profiling over
the nuisance parameters.  Correlations between observables like
different channels at the LHC are taken into account. These likelihood
maps can be projected on two-dimensional planes as well as
single-parameter measurements using profile likelihood or Bayesian
methods. Throughout this paper we will apply frequentist profile
likelihoods and this way avoid volume effects.

For the TeV-scale fits the same technique is used to select points of
the Markov chains following the proposal function.  When we display
observables instead of the likelihood function, only points that pass
sharp criteria, \eg $\Omega_{\lsp} h^2$ within the theory uncertainties,
are displayed. Similar to a profile likelihood, we assign the value of
the point with highest likelihood to each bin. \bigskip

For constraints specifically for the NMSSM we rely on
\textsc{NMSSMTools}~\cite{nmssmtools} to calculate the supersymmetric
mass spectrum, the Higgs branching ratios, the $B$ observables,
$(g-2)_\mu$, and electroweak precision observables. The relic density
and the direct detection limits are calculated with
\textsc{MicrOMEGAs}~\cite{micromegas}. The number of events in the
Higgs production channels at the LHC for the SM-like $H_{125}$ are
computed using the standard \textsc{SFitter}-Higgs
setup~\cite{sfitter_higgs}, \textsc{HDecay}~\cite{s-hit}, and
\textsc{NMSSMTools}.

For the MSSM study in the appendix the supersymmetric spectrum is
calculated using \textsc{SuSpect3}~\cite{suspect} while the Higgs
branching ratios are computed using \textsc{Susy-Hit}\cite{s-hit} and
\textsc{HDecay}.  For the electroweak precision observables we rely on
\textsc{SusyPope}~\cite{Heinemeyer:2007bw}. Finally, we use
\textsc{SuSpect3} to compute the $B$-observables and $(g - 2)_\mu$. As
for the NMSSM the relic density and the direct detection limits are
calculated with \textsc{MicrOMEGAs}.

%%%%%%%%%%%%%%%%%%%%%%%%%%%%%%%%%%%%%%%%%%%%%%%%%%%%%%%%%%%%%%%%%%%%%%%%%%%%%%%%
\section{TeV-scale NMSSM}
\label{sec:low_scale}

The NMSSM can explain or motivate interesting measurements which are
not accessible in the MSSM. This is typically an effect of the
extended Higgs particle content, for example including a second
pseudoscalar state. One of these measurements is the galactic center
excess in gamma rays, based on light neutralinos and a light
pseudoscalar mediator. Another is invisible decays of the SM-like
Higgs boson. It will turn out that both of them are strongly linked.

We assume a $\mathbb{Z}_3$-symmetric NMSSM, where all input parameters
are set at the scale 1~TeV. The absence of unifying
assumptions leads to a large number of model parameters, namely the
slepton and squark masses, the trilinear couplings and the gaugino
mass parameters. In this study we decouple the squarks, sleptons, and
gluinos by setting the soft masses to 10~TeV and the trilinear
couplings to zero, because these sectors are not experimentally
relevant.  To obtain the correct 125~GeV Higgs mass we adjust the stop
masses in the TeV-range appropriately. For the Higgs--singlet sector
the relevant input parameters are
\begin{align}
\{ \lambda, \tkappa, \Alambda, \Akappa, \tan \beta, \mu \} \; .
\end{align}
Alternatively, we can replace $A_{\lambda,\kappa}$ by the diagonal
entries in the pseudoscalar mass matrix of
Eq.\eqref{eq:pseudo_matrix}. The neutralino--chargino adds the free
parameters $M_1$ and $M_2$.\bigskip

While $\Akappa$ is given at the SUSY scale of 1~TeV, NMSSMTools
computes the Higgs masses at the averaged squark masses, in our case
around 10~TeV. In Sec.~\ref{sec:high_scale} we will discuss the scale
dependence of the singlet related parameters in detail. The
approximate form of $\Akappa$ at the averaged squark mass scale can be
computed from the RGE via
\begin{align}
A_{\kappa}(10 ~\tev) &= A_{\kappa,0} 10^{\frac{3\lambda \tkappa^2}{4 \pi^2}}+ \left( 10^{\frac{3\lambda \tkappa^2}{4 \pi^2}}-1\right) \frac{\Alambda}{\tkappa^2} \notag \\
&\approx A_{\kappa,0} +  \frac{3  \log 10}{4 \pi^2} \lambda \Alambda \; .
\label{eq:akappa}
\end{align}
Due to the large value of $|\Alambda|$, that further increases with
the absolute value of $\mu$, $\Akappa$ increases from $-250
\unit{GeV}$ to the order of $30 \unit{GeV}$ at 10 TeV for
$\lambda=0.2$ and $\Alambda=8 \unit{TeV}$. When we consider small
singlet masses, the scale dependence of $\Akappa$ plays an important
role in their parameter dependence.

%%%%%%%%%%%%%%%%%%%%%%%%%%%%%%%%%%%%%%%%%%%%%%%%%%%%%%%%%%%%%%%%%%%%%%%%%%%%%%%%
\subsection{Galactic center excess}
\label{sec:hooperon}

With the galactic center excess of gamma ray photons we add an
experimental motivation for a light neutralino in combination with a
light pseudoscalar mediator~\cite{hooperon,hooperon_simplified} to
the (largely negative) experimental results in Tab~\ref{tab:data}. The
spherically symmetric excess can be explained by annihilating dark
matter, more specifically by a neutralino in the range of 30~GeV to
40~GeV or even 70~GeV, that annihilates into a pair of fermions, for
example $\lsp \lsp \to b \bar{b}$. For a type-II two-Higgs doublet
model the bottom and tau Yukawa couplings are aligned, which means
that the assumed decay to a $b\bar{b}$ pair will dominate over the
accompanying decay to $\tau^+ \tau^-$. In our preferred data regions
the $A_1$ decays into $b \bar{b}$ (94\%) as well as into $\tau
\bar{\tau}$ (6\%) pairs, the ratio reflecting the size of the Yukawa
couplings and the color factor in the case of quarks.

In the MSSM such a light neutralino LSP is a problem, because it will
lead to too large a relic density. In the presence of direct SUSY and
Higgs search results there is no obvious way to annihilate it
efficiently enough to arrive at the observed relic density. Different
co-annihilation
channels~\cite{stau-co-annihilation,char-co-annihilation,stop-co-annihilation}
require new charged states in reach of LEP2 and are therefore not
viable~\cite{lep_constraints}.

In contrast, the two pseudoscalars in the NMSSM can mediate a sufficiently
fast annihilation, because the LSP annihilation through its resonance
pole is not $p$-wave suppressed like it is for vector bosons or
scalars. While the mass of the MSSM-like pseudoscalar Higgs is
strongly constrained by heavy neutral and charged Higgs searches, the
additional light pseudoscalar can be mostly singlet. The neutralino
coupling to such a mediator is given in Eq.\eqref{eq:achichi} for a
largely singlino mediator turning into Eq.\eqref{eq:a_coup}.\bigskip

We consider a generic NMSSM scenario based on a light singlino with a
bino and a Higgsino admixture, \ie $\tkappa \ll
1$~\cite{hooperon_felix}. An alternative solution presented in
Ref.~\cite{papucci_zurek} combines a bino-Higgsino LSP with an NMSSM
pseudoscalar mediator, but will not lead to the measurable invisible
Higgs decay we are interested in.  Light neutralinos with a sizeable
wino or Higgsino component are essentially ruled out by $Z$-pole
measurements and by chargino searches at LEP, so we decouple the wino
at $M_2=3$~TeV.  In this limit we can link $\tkappa$ to the neutralino
mass through Eq.\eqref{eq:mass_approx}.  We then adjust $\Alambda$
such that the singlet component of the SM-like Higgs is minimized,
avoiding Higgs sector constraints altogether~\cite{sfitter_higgs}.

If the annihilation process leading to the galactic center excess
proceeds via a pseudoscalar decaying into a pair of bottom quarks,
today's dark matter annihilation cross section
is~\cite{hooperon_simplified}
\begin{align}
\sigma v \Big|_{v = 0} 
\approx 
  \dfrac{3}{2\pi} \, 
  \sqrt{ 1 - \dfrac{m_b^2}{m_{\lsp}^2} } \, 
  \dfrac{g_{{A_1}\lsp\lsp}^2  g_{{A_1} b b}^2  m_{\lsp}^2}
  {\left(m_{A_1}^2-4 m_{\lsp}^2\right)^2
  + m_{A_1}^2 \Gamma_{A_1}^2} \; .
\end{align}
Large $\tan \beta$ increases the coupling to the
bottom quarks given in Eq.\eqref{eq:abb}. A strong coupling to the
bottom quarks also leads to a small width of the light pseudoscalar
which further increases the cross section when one is close to the
on-shell condition.\bigskip

%----------------------------------------------------
\begin{figure}[t]
 \includegraphics[width=0.48\textwidth]{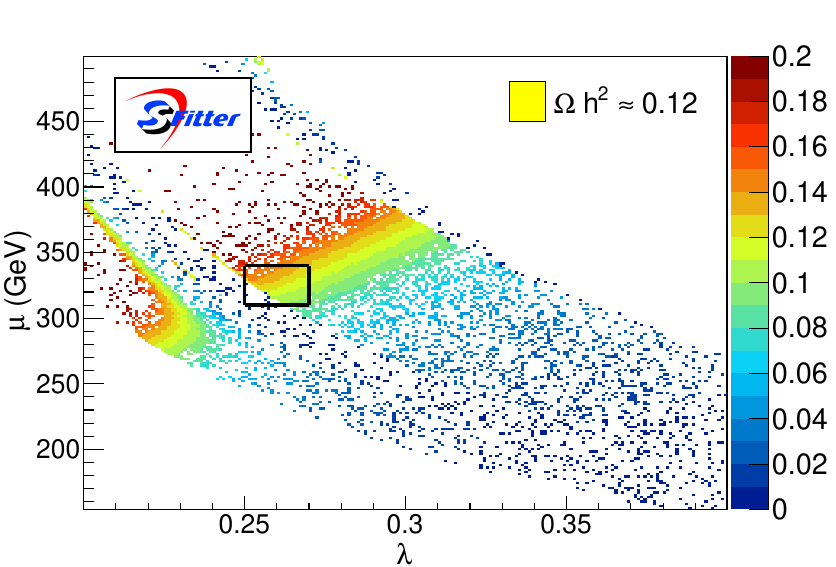}
  \includegraphics[width=0.48\textwidth]{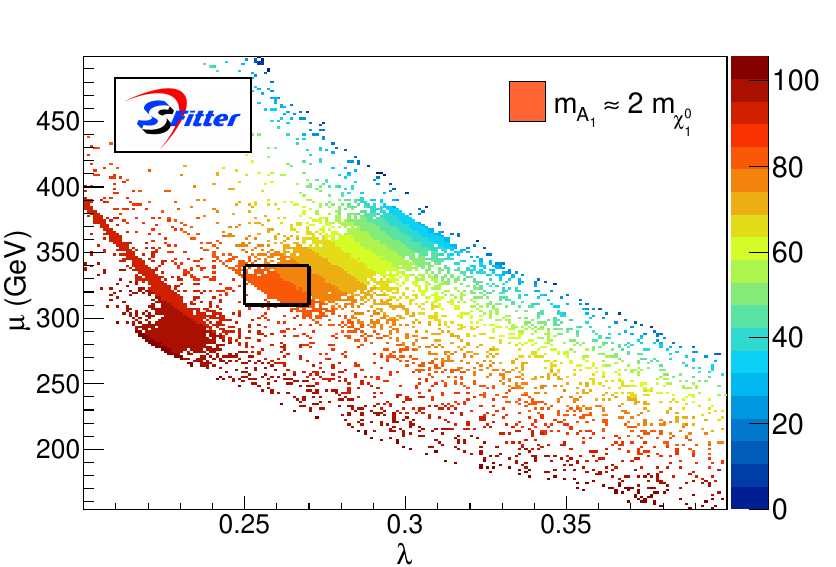} \\
  \includegraphics[width=0.48\textwidth]{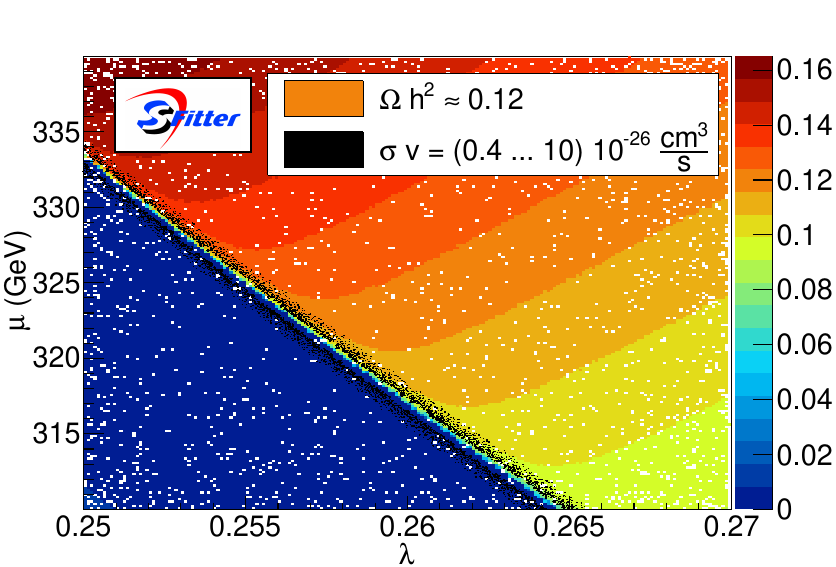}
  \includegraphics[width=0.48\textwidth]{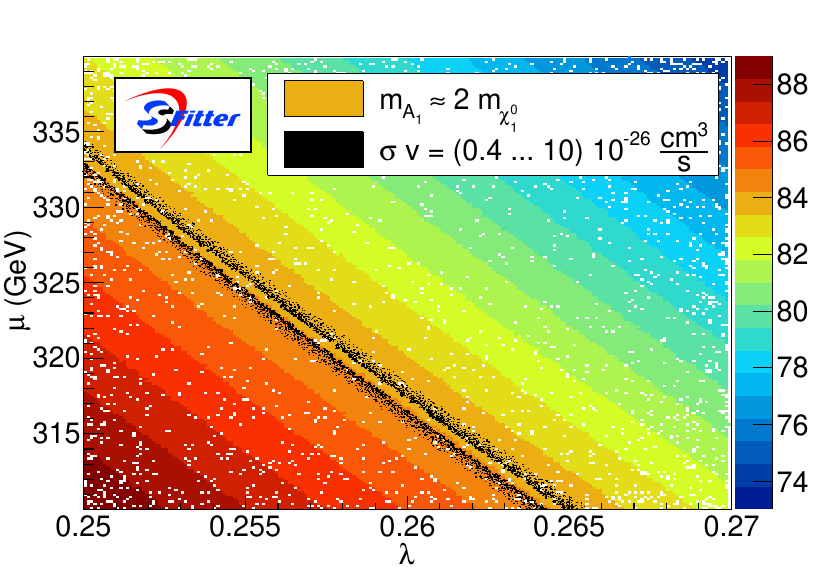}
  \caption{Neutralino relic density (left) and mass of the light
    pseudoscalar $A_1$ (right) color-coded as a function of $\mu$ and
    $\lambda$. The two lower panels are zoomed into the respective
    upper panels. The orange regions in the lower left panel are
    compatible with the relic density $\Omega_{\lsp} h^2 = 0.107~...~0.131$,
    considering the theoretical uncertainty. In addition to the usual
    decoupling through large scalar masses we fix $\tan \beta=40$,
    $\Akappa=-250$~GeV, $\Alambda$ according to the decoupling
    condition Eq.\eqref{eq:decoup_higgs}, and $\tkappa$
    corresponding to an LSP mass of 40~GeV through
    Eq.\eqref{eq:mass_approx}.}
  \label{fig:GCE} 
\end{figure}
%----------------------------------------------------

Using \textsc{Sfitter} we analyze the NMSSM parameter space with a
focus on the SM-like Higgs mass, the LSP mass, the observed relic
density $\Omega_{\lsp} h^2$, the $Z$ width measurements, the Xenon
direct detection constraints, and the galactic center excess. As
expected from Sec.~\ref{sec:higgs_singlet} the key model parameters
are the Higgsino mass parameter $\mu$, the singlet mass parameter
$\tkappa$, and the coupling $\lambda$, which links the singlet to the
MSSM Higgs sector. In Fig.~\ref{fig:GCE} we observe a broad band in
the $\lambda$ vs $\mu$ plane, which is defined by non-zero values for
the Higgs masses: for large $\mu \gg m_Z$ and $\lambda \to 1$ the mass
of the lightest CP-odd scalar $A_1$ vanishes. This follows from
Eq.\eqref{eq:akappa} and the leading term in
Eq.\eqref{eq:mass_approx}, such that the pseudoscalar mass matrix in
Eq.\eqref{eq:pseudo_matrix} only depends on $\mu$ and $\lambda$.
$\Akappa$ is of the order of 100 GeV at the scale of the averaged
squark masses.  Following Eq.\eqref{eq:akappa}, an increase of
$\lambda$ and $\Alambda$, leads to an increase of $\Akappa$. This
results in a smaller mass eigenvalue of $A_1$. As $\Alambda$
increases with $\mu$ following Eq.\eqref{eq:decoup_higgs}, higher
values of $\mu$ lead to smaller pseudoscalar mass.

The mass of the lightest CP-even scalar $H_1$ is given by the $(3,3)$
entry of the Higgs mass matrix in Eq.\eqref{eq:higgs_matrix} in the
Higgs decoupling limit Eq.\eqref{eq:decoup_higgs},
\begin{align}
M_{H,h,S}^2 \Bigg|_{33}
&=m_Z^2 \left(
    \dfrac{\lambda^2}{g^2}  \left( 1 - s_{2\beta} \tkappa \right)
  + \dfrac{\tkappa  \mu}{m_Z^2} 
      \left(\Akappa+4\tkappa \mu\right)
\right) \; .
%\notag \\
%&=m_Z^2 \left(
%    \dfrac{\lambda^2}{g^2}  \left( 1 - s_{2\beta} \dfrac{m_{\lsp}}{2\mu} \right)
%  + \dfrac{m_{\lsp}}{2m_Z^2} 
%      \left(\Akappa+2m_{\lsp}\right)
%\right) 
\end{align}
$\Akappa$ enters here with a positive sign, so that for small $\mu
\lesssim m_Z$ and $\lambda \to 0$ the mass of this CP-even singlet
vanishes. Possible experimental constraints are expected to further
reduce this band.\bigskip

Within this broad band shown in the upper panels of Fig.~\ref{fig:GCE}
the structure originates from two sets of input parameters to the
calculation of the relic density. On the one hand there is a strong
dependence on the mass of the pseudoscalar mediator, on the other
hand the couplings of the LSP depend on the gaugino and Higgsino
content of the LSP.

In the upper part of the band, with $m_{A_1}<80$~GeV and
$0.26<\lambda<0.3$ the dark matter annihilation is mediated by the
$Z$-funnel, with a coupling to the Higgsino content proportional to
$N_{13}^2-N_{14}^2$ as given in Eq.\eqref{eq:z_coup}. Smaller values
of $\lambda$, correlated with large values of $\mu$, decouple the
singlet/singlino sector from the MSSM. An efficient dark matter
annihilation is not possible, and the relic density is too large. On
the other hand, too small values of $\mu$ and large $\lambda$ increase
the Higgsino--singlino mixing via the off-diagonal terms in the
neutralino mass matrix. The Higgsino component in the relic neutralino
then results in too small a relic density. In between,
Fig.~\ref{fig:GCE} indicates a well-defined regime with the correct
relic density. The corresponding mass of the lightest pseudoscalar
$A_1$ indicates that the resonance condition $m_{A_1} \approx 2 m_{\lsp}$
is only fulfilled at the lower end of this regime, while the larger
part of the allowed band relies on $Z$-mediated dark matter annihilation.

In the lower part of the band, defined by the onset of the resonance
condition $m_{A_1} \approx 2 m_{\lsp}$, a steep
decrease of the relic density leaves a very narrow strip where
the annihilation proceeds via the $A$-funnel and reproduces the
correct value of $\Omega_{\lsp} h^2$. With the increasing $A$-funnel
contribution, the $Z$-mediated annihilation rate has to decrease,
which means that the allowed region bends towards larger values of
$\mu$.  The moment the resonance condition is actually fulfilled, the
annihilation through the $A$-funnel becomes too efficient, and the
predicted relic density drops well below the measured value.  For
$m_{A_1} \approx 90$~GeV, corresponding $\lambda<0.24$ and $\mu\approx
300$~GeV, the annihilation again proceeds off-shell, predicting the
correct relic density starting with a reduced $Z$-mediated rate at
large $\mu$.  At $\lambda \approx 0.225$ and $\mu \approx 275$~GeV the
annihilation is again mediated by the $Z$-boson alone.

A few hardly visible points with the correct relic density at the very
top of the allowed mass band arise from $H_1$-funnel annihilation and
will be of no relevance to our further discussion, because the scalar
mediator with its $p$-wave suppression fails to explain
the galactic center excess.\bigskip

The lower panels of Fig.~\ref{fig:GCE} focus on $A_1$-funnel
annihilation just below the resonance condition. There are two regions
where the annihilation cross section is compatible with the galactic
center excess --- within a comparably broad range of $\sigma v =
(0.4~...~10) \cdot 10^{-26} \cmu^3/\text{s}$. The region below the
resonance condition $m_{A_1} \lesssim 2 m_{\lsp}$ is compatible with
the relic density, while the other one is not.

%%%%%%%%%%%%%%%%%%%%%%%%%%%%%%%%%%%%%%%%%%%%%%%%%%%%%%%%%%%%%%%%%%%%%%%%%%%%%%%%
\subsection{Invisible Higgs decays}
\label{sec:hinv}

Invisible Higgs decays have long been in the focus of LHC
studies~\cite{eboli_zeppenfeld,jamie}. At the LHC the upper limits on
an invisible branching ratio are 57\% in the WBF channel from
CMS~\cite{hinv_cms} and 78\% combining associated Higgs production
and gluon fusion from ATLAS~\cite{hinv_atlas}. Eventually, the HL-LHC
should be sensitive to invisible branching ratios of a few per-cent in
the WBF production channel~\cite{jamie}. Usually, such invisible Higgs
decays are associated for example with a Higgs portal to a scalar dark
matter sector~\cite{portal}. We will show that in the NMSSM, invisible
Higgs searches can also probe a relevant part of the
dark-matter-related parameter space through the decay $H_{125} \rightarrow
\lsp \lsp$. Because this decay requires relatively light LSP
neutralinos these scenarios can be linked to the galactic center
excess discussed in Sec.~\ref{sec:hooperon}.\bigskip

The decay width of the CP-even Higgs into two neutralinos is given by
\begin{align}
\Gamma (H_{125} \rightarrow \lsp \lsp)= 
\dfrac{m_{H_{125}}}{16 \pi} \;
g_{H_{125} \lsp \lsp}^2 \;
\Lambda^{3/2}\; , 
\qqquad \text{with} \quad 
\Lambda =1-\dfrac{4 m_{\lsp}^2}{m_H^2}  \; .
\end{align}
The Higgs--LSP coupling in the MSSM is
\begin{align}
g_{H \lsp \lsp} \Bigg|_\text{MSSM}
= & \left( g N_{11}-g' N_{12} \right) \; 
  \left( \sin \alpha N_{13} + \cos \alpha N_{14} \right) \notag \\
\equiv & \left( g N_{11}-g' N_{12} \right) \; 
  \left( S_{21} N_{13} + S_{22} N_{14} \right) \; .
\label{eq:gMSSM}
\end{align}
The $N _{1j}$ are the entries of the neutralino mixing matrix, and
$S_{2i}$ are the entries of the CP-even Higgs mixing matrix. In the
simple $(2 \times 2)$ case the latter are expressed in terms of the
mixing angle $\alpha$. In the MSSM invisible Higgs decays have to be
mediated by gaugino and Higgsino fractions combined, or more
specifically by a mixed bino--Higgsino LSP. In the NMSSM this coupling
receives additional contributions from the singlet, namely
\begin{align}
g_{H \lsp \lsp}
= g_{H \lsp \lsp} \Bigg|_\text{MSSM}
+ \sqrt{2}\lambda \left[ (S_{21} N_{14} + S_{22} N_{13}) N_{15} 
                        + S_{23} N_{13} N_{14} 
- \tkappa S_{23} N_{15}^2 \right]  \; .
\label{eq:higgs_coup}
\end{align}
Now, invisible Higgs decays can proceed to bino--Higgsino,
singlino--Higgsino, and in the presence of a singlet component in the
Higgs boson even pure Higgsino and pure singlino LSPs.\bigskip

Just like the galactic center excess, invisible decays of the SM-like
Higgs benefit from a light, mixed neutralino LSP.  Decoupling squarks,
sleptons, and gluino we can ask if there are regions with a large
branching ratio $H_{125} \rightarrow \lsp \lsp$ for a mixed
bino--Higgsino--singlino LSP, and such parameter regions can be
related to the galactic center excess.

Following Eq.\eqref{eq:mass_approx} the mass of the singlino LSP is
given by $2 \mu \tkappa$. To suppress the Higgsino component we
require $\tkappa < 1/2$ and keep $\mu > 190$~GeV in a first
step. This way the Higgsino component in the LSP ranges around 5\% to
10\%, leading to a sizeable coupling to the Higgs sector but
preventing a large coupling to the $Z$-boson.  Our choice of
parameters for the galactic center excess fixes $\tan \beta= 40$ and
$\Akappa < -250$~GeV. The coupling condition
Eq.\eqref{eq:decoup_higgs} determines $\Alambda$ to minimize the
singlet component of the SM-like Higgs. The remaining parameters are
$\lambda, \tkappa$ and $M_1$.  Small values of $M_1$ increase
the bino component of the LSP while small values of $\tkappa$
increase the singlino component. To keep a constant hooperon mass,
$\tkappa$ and $M_1$ have to behave inversely
proportional.\bigskip

%----------------------------------------------------
\begin{figure}[t]
  \includegraphics[width=0.32\textwidth]{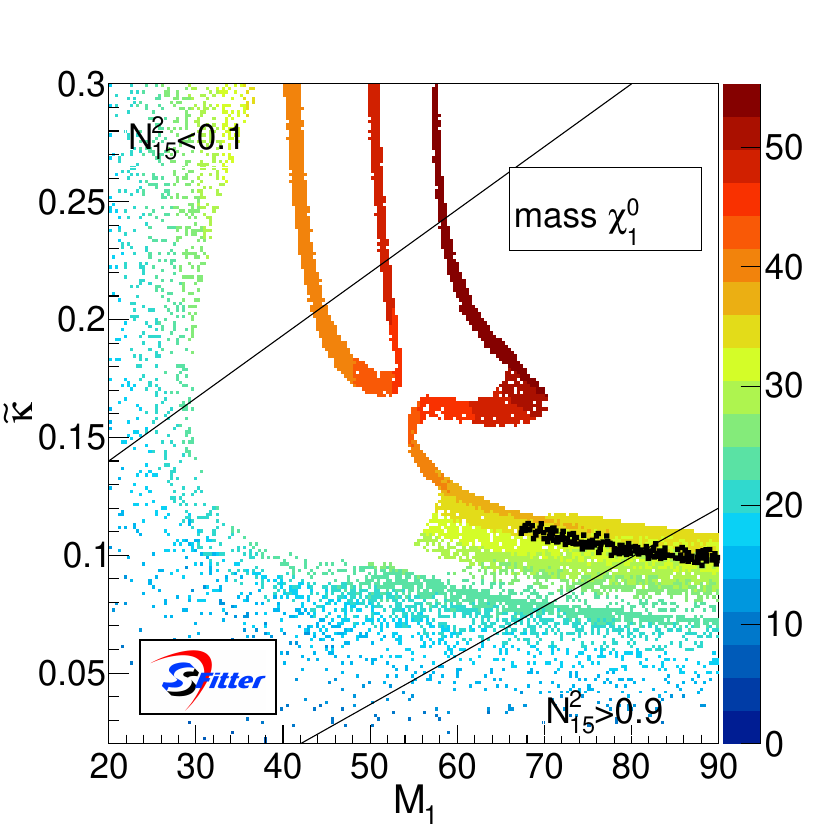} 
  \includegraphics[width=0.32\textwidth]{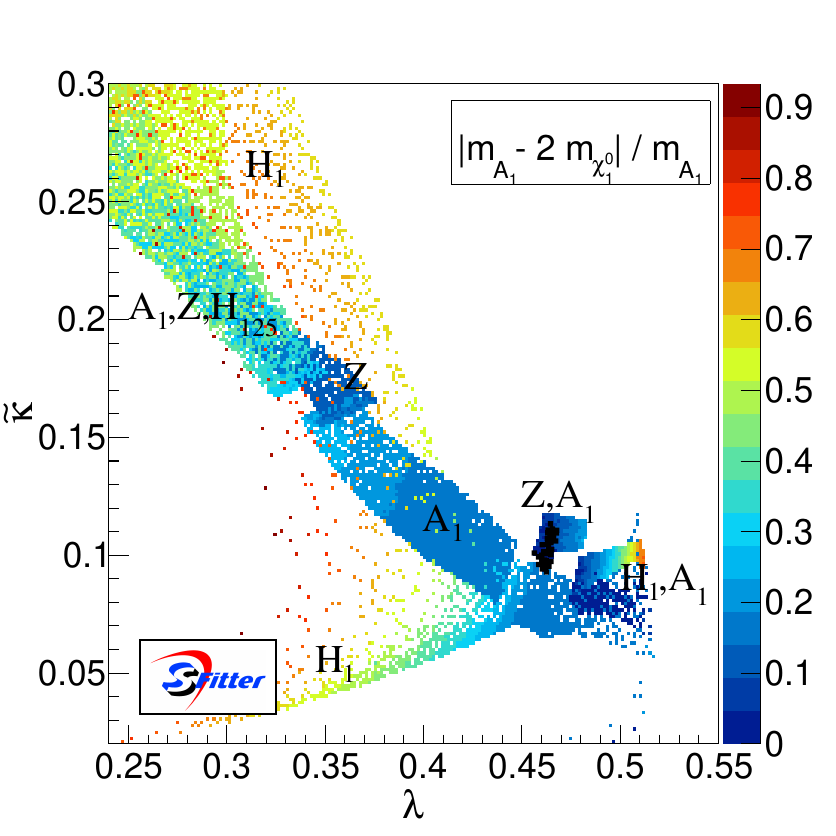} 
  \includegraphics[width=0.32\textwidth]{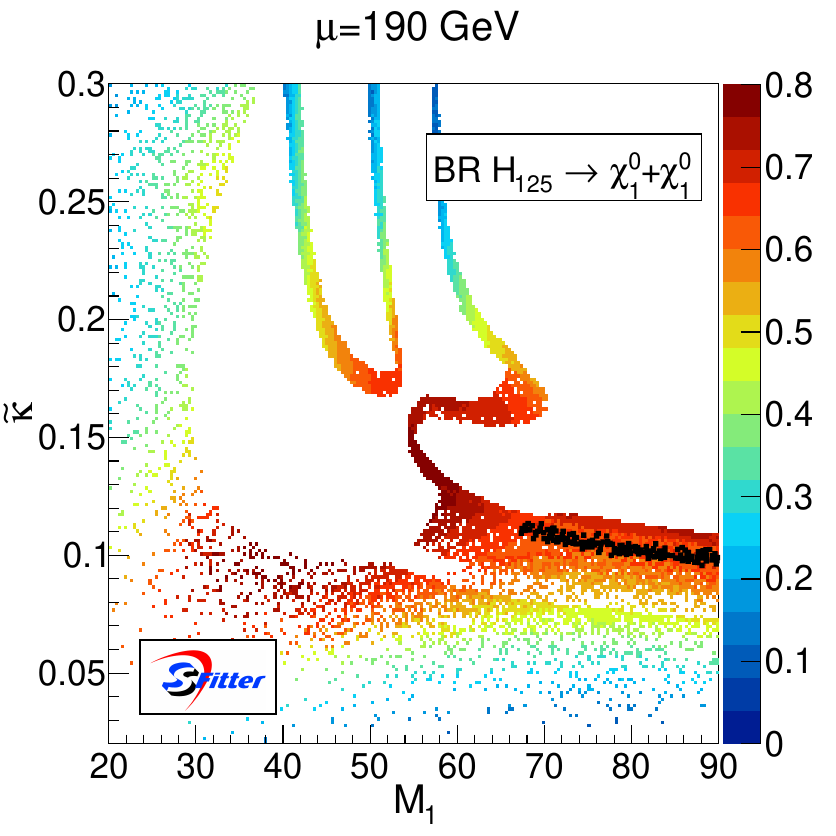} 
  \caption{LSP mass, mass splitting between the LSP and the
    pseudoscalar mediator, and invisible Higgs branching ratio with
    $\mu=190$~GeV and $\Akappa = -250$~GeV in the $\tkappa -
    M_1$, $\tkappa - \lambda$, and $\tkappa - M_1$ planes.
    All displayed points are compatible with the relic density, Xenon,
    a chargino mass above 103~GeV, and the correct SM-like Higgs
    mass. Moreover, they always have an invisible branching ratio $\br
    (H_{125} \rightarrow \lsp \lsp) > 10\%$. The black points are
    consistent with the galactic center excess.}
  \label{fig:BR_190} 
\end{figure}
%----------------------------------------------------

In Fig.~\ref{fig:BR_190} we show the result of the \textsc{SFitter}
analysis, starting with fixed $\mu = 190$~GeV.  Of the experimental
constraints discussed in Sec.~\ref{sec:data} we now only include the
relic density, the Xenon constraints, the chargino mass constraints,
the invisible $Z$-width constraint $\Gamma_{Z \rightarrow
  \text{inv}}<2~\mev$ and the SM-like Higgs mass constraint $m_H =
122~...~128$~GeV.  We only show parameter points with $\br (H_{125}
\rightarrow \lsp \lsp) > 10\%$. For our starting value $\mu = 190$~GeV
we fix $\Akappa = -250$~GeV.

Three distinct, narrow strips for example in the upper row of
Fig.~\ref{fig:BR_190} are defined by constant LSP masses around
40~GeV, 48~GeV, and 55~GeV. For $m_{\lsp} \approx 55$~GeV the
annihilation is mediated by $H_{125}$. If the mass moves closer to the
$H_{125}$ on-shell condition the relic density becomes too small. At
the lower end of the 55~GeV strip the additional annihilation through
the $A_1$ pseudoscalar and the $Z$ becomes too strong to reproduce the
observed relic density. Numerically, the reason is that $m_{A_1}
\propto \tkappa$ reaches 120~GeV around $\tkappa =0.16$,
which opens a pseudoscalar-mediated LSP annihilation channel. For
$\mu=190$~GeV this coincides with the possibility to efficiently
annihilate though an $s$-channel $Z$-exchange.

The 40~GeV and 48~GeV strips are defined by the $Z$-mediated LSP
annihilation. Each of them lives on one side of the $Z$-pole, because
the annihilation on the pole is too efficient to give the correct
relic density. Both strips follow the asymptotic behavior of the LSP
mass. The 40~GeV strip continues towards high values of $M_1$, but
with a reduced LSP mass. The reason is that the additional
annihilation mediated by $A_1$ adds to $Z$-mediated
annihilation. Finally, the annihilation via the pseudoscalar connects
the annihilation channels for $m_{\lsp}=48$ GeV and $m_{\lsp}=55$ GeV
around $M_1=70$.

In the broad regions with $M_1= 10~...~40$~GeV the annihilation is in
addition mediated by $H_1$ ($\tkappa > 0.1$) and by a combination of
$H_1$ and $A_1$ ($\tkappa < 0.1$).

One key feature is the hole in the allowed parameter space
around $\tkappa = 0.15$ and $M_1 = 50$~GeV.  For example along
the 40~GeV strip the LSP composition changes from bino to singlino,
with a 5\% to 12\% Higgsino contribution.  While the sum of the two
Higgsino components increases towards the singlino LSP, their ratio
switches. This leads to an intermediate region where both components
have a similar value.  At this point $g_{Z \lsp \lsp}$ given by
Eq.\eqref{eq:z_coup} vanishes, interrupting the $Z$-mediated
annihilation. The other annihilation channels are weak, so the relic
density is too large.\bigskip

In the right panel of Fig.~\ref{fig:BR_190} we see that the parameter
points which can account for the galactic center excess appear
precisely where we also expect invisible decays for the SM-like Higgs
$H_{125}$.  Indeed, for $M_1=50~...~60$~GeV and $\tkappa \approx
0.1$ we find the highest branching ratio $H_{125} \rightarrow \lsp
\lsp$. This region has unique properties: as discussed above, the dark
matter annihilation proceeds through an on-shell $Z$-boson, with
$m_{\lsp} = 45\pm 2$~GeV. The LSP is a mixed state with 8\% Higgsino,
30\% to 50\% singlino, and 40\% to 50\% bino content. The two Higgsino
components are of the same size, strongly reducing the $Z$-coupling
$g_{Z \lsp \lsp}$.  The lightest pseudoscalar $A_1$ has a mass around
115 to 135~GeV, but as an at least 95\% singlet remains undetected. 
Because of the strongly mixed LSP content, each $N_{1i}$ term can
contribute to $g_{H_2 \lsp \lsp}$. This large value induces an
invisible branching ratio of up to\footnote{In this discussion we
  ignore the limits on invisible Higgs decays for example from the
  \textsc{SFitter} Higgs analysis~\cite{legacy}.  There, we find
  $\br_\text{inv}<30.6\%$ at 95\% C.L. in an 8-parameter coupling
  analysis. For a dedicated NMSSM fit this limit is expected to be
  even more constraining.}.
\begin{align}
\br(H_{125}\rightarrow \lsp \lsp) \lesssim 70\% 
\qqquad \text{for} \quad \mu = 190~\gev \; ,
\end{align}
for the parameter space consistent with the galactic center excess.

In this region with an overwhelming invisible Higgs decay width we
need check a few experimental constraints not represented in
Fig.~\ref{fig:BR_190}: First, the cancellation of the two Higgsino
components renders the partial decay width $\Gamma_{Z\rightarrow \text{inv}}$
smaller than 0.1~MeV. For a mostly-singlino LSP region it increases
through large $N_{14}$ values up to 2~MeV.  The Xenon limits
on direct dark matter detection exclude the region centered around
$M_1=40$~GeV and $\tkappa=0.2$, linked to the $H_1$ and $A_1$
annihilation channels. However, none of these additional constraints
affect the parameter regions with invisible Higgs branching ratios
between 10\% and 30\%.\bigskip

%----------------------------------------------------
\begin{figure}[t]
  \includegraphics[width=0.32\textwidth]{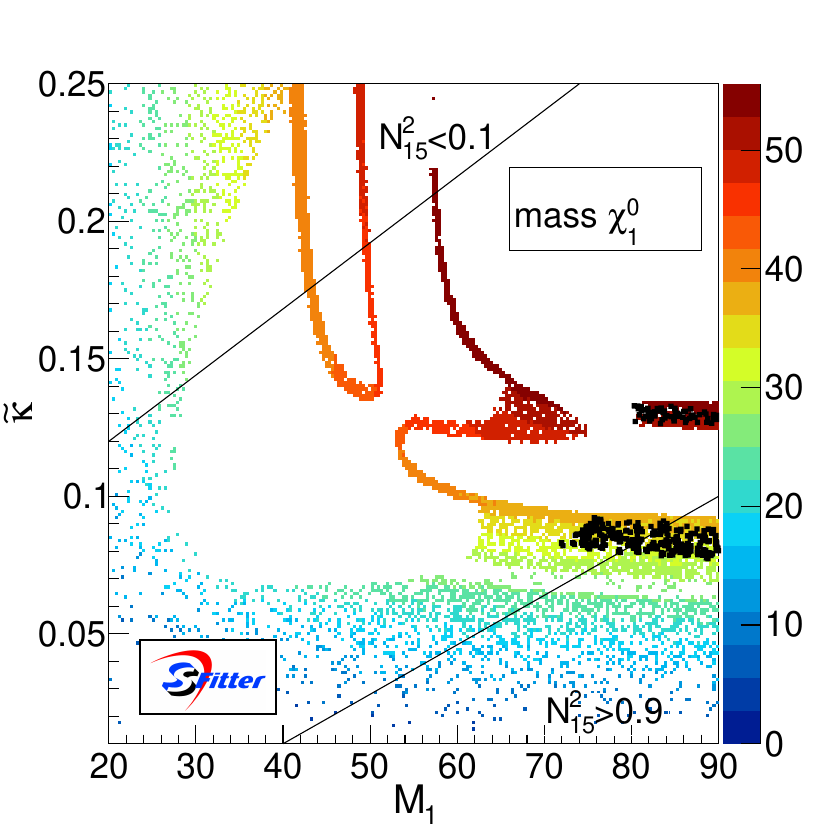} 
  \includegraphics[width=0.32\textwidth]{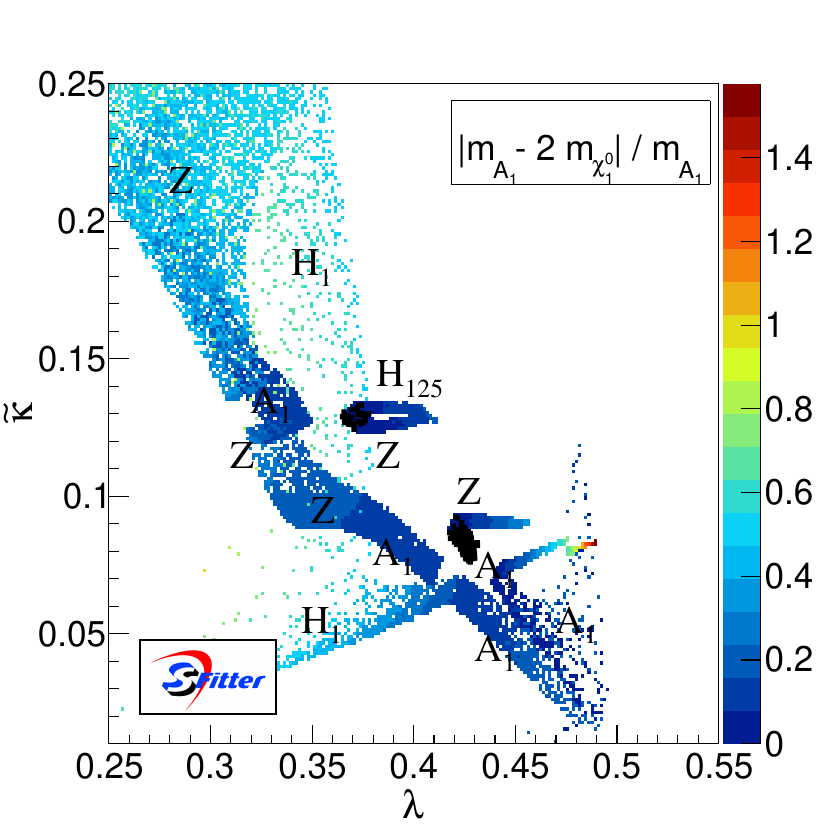} 
  \includegraphics[width=0.32\textwidth]{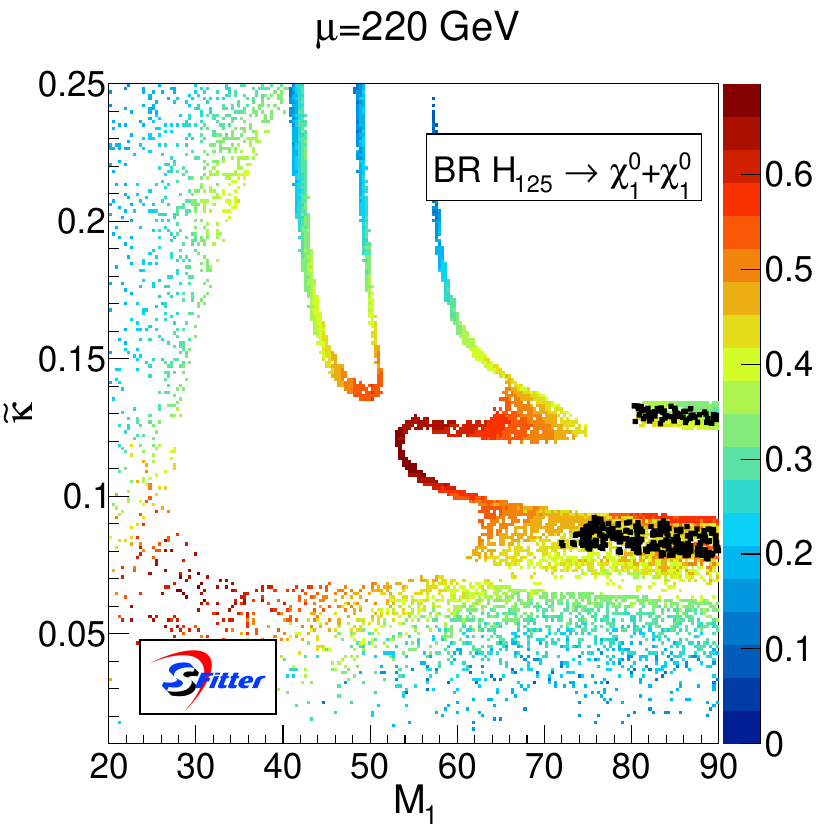} \\
  \includegraphics[width=0.32\textwidth]{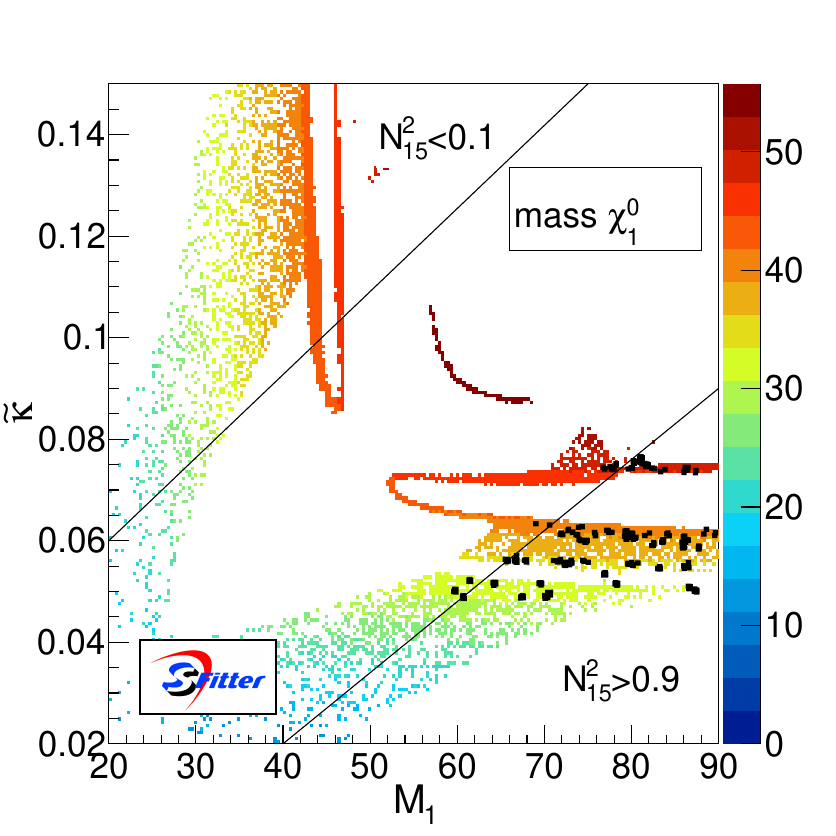} 
  \includegraphics[width=0.32\textwidth]{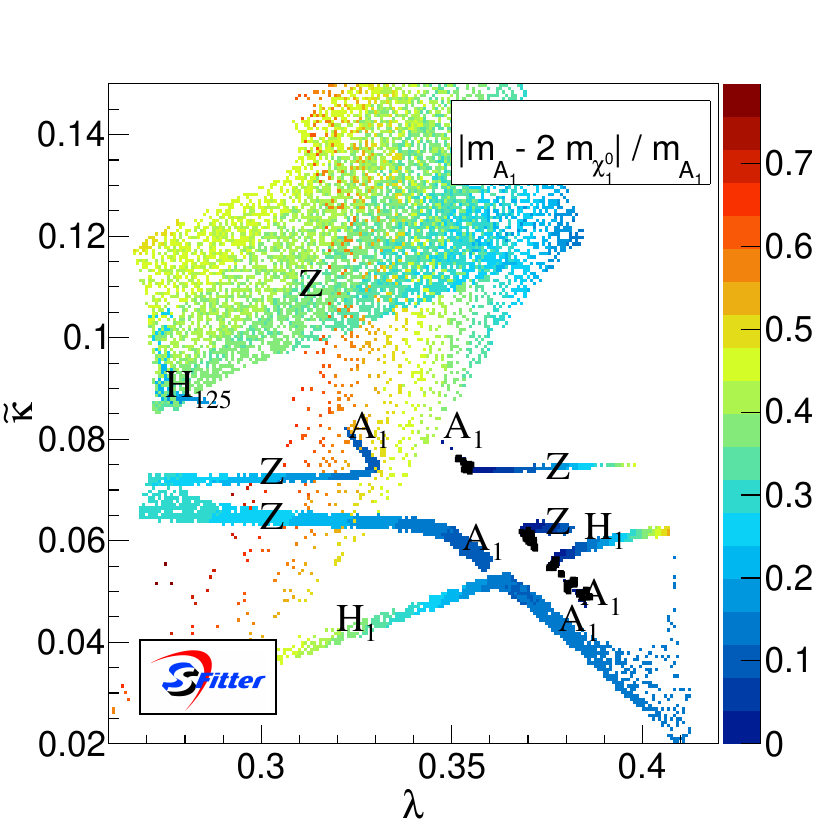}
  \includegraphics[width=0.32\textwidth]{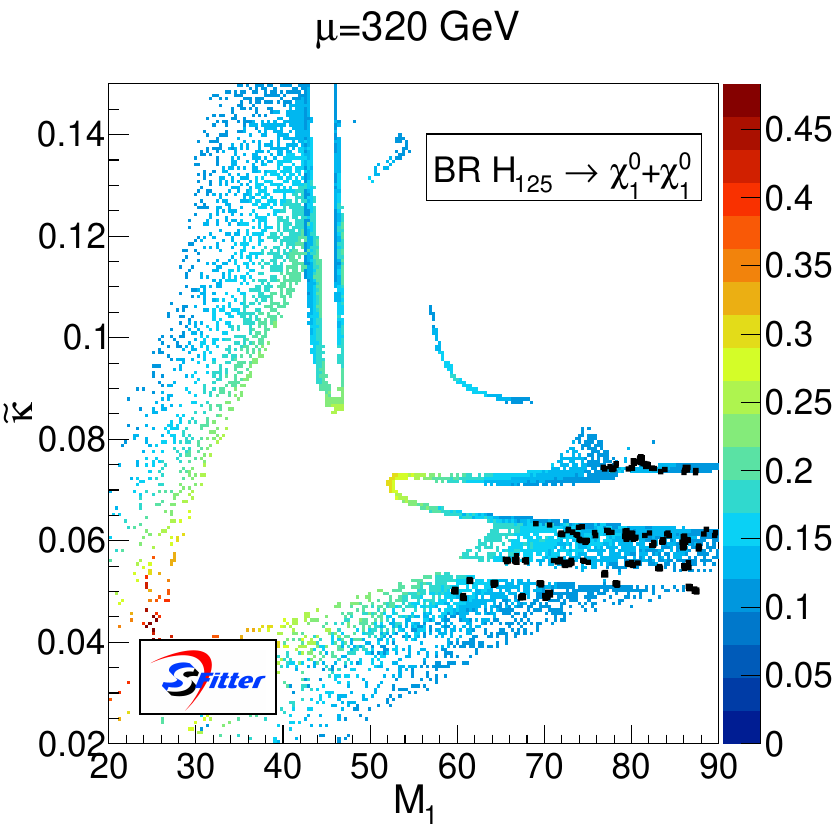}
  \caption{LSP mass, mass splitting between the LSP and the
    pseudoscalar mediator, and invisible Higgs branching ratio now for
    $\mu=220$~GeV (upper row) and $\mu = 320$~GeV (lower row) in the
    $\tkappa - M_1$, $\tkappa - \lambda$, and $\tkappa - M_1$
    plane. Correspondingly we choose $\Akappa = -280$~GeV and
    -380~GeV. As in Fig.~\ref{fig:BR_190} all displayed points are
    compatible with the relic density, Xenon, a chargino mass above
    103~GeV, and the correct SM-like Higgs mass. Moreover, they always
    have an invisible branching ratio $\br (H_{125} \rightarrow \lsp
    \lsp) > 10\%$. The black points are consistent with the galactic
    center excess.}
  \label{fig:BR_all} 
\end{figure}
%----------------------------------------------------

In the next step we vary $\mu$, and with it $\Akappa$ to keep the singlet 
Higgs masses in the desired range. This means that for $\mu=220$~GeV and
$\mu=320$~GeV we have to set $\Akappa= -280$~GeV and $\Akappa=
-350$~GeV, respectively.

Increasing $\mu$ slowly impacts the LSP annihilation channels.  First,
we see that the allowed regions for $\mu=220$~GeV shown in the upper
left panel of Fig.~\ref{fig:BR_all} are very similar the case of
$\mu=190$~GeV. This indicates that our above results are not very
fine-tuned. For a fixed LSP mass an increase of $\mu$ merely leads to
a smaller Higgsino component, which in turn leads to a smaller $Z\lsp
\lsp$ coupling. It also decreases the invisible Higgs branching to
\begin{align}
\br(H_{125}\rightarrow \lsp \lsp) \lesssim  40\% 
\qqquad \text{for} \quad \mu = 220~\gev \; ,
\end{align}
in the relevant parameter region for the galactic center
excess.\bigskip

For larger $\mu=320$~GeV the picture changes: for a fixed LSP mass an
increase of $\mu$ leads to a smaller Higgsino content.  While for
$\mu=190$~GeV the two Higgsino components add to $5\%~...~17\%$, they
now stay in the $2\%~...~5\%$ range.  This immediately leads to a
smaller $Z\lsp \lsp$ coupling --- and a reduced invisible Higgs
branching ratio.  The $Z\lsp \lsp$ coupling implies that the LSP mass
has to be closer to the on-shell condition to give the correct relic
density. Indeed, for a bino-like LSP and $\mu=320$~GeV, the lowest
mass strip is now defined by $m_{\lsp} \approx 42$~GeV instead of
40~GeV. Similarly, the high-mass strip moves down from $m_{\lsp} =
48$~GeV to 46~GeV. Altogether, in the lower left panel of
Fig.~\ref{fig:BR_all} we see that the annihilation regions mediated by
the $Z$-funnel and the $H_1$ funnel with $m_{\lsp} = 55$~GeV clearly
separate.\bigskip

The annihilation processes can now best be identified in the central
lower panel of Fig.~\ref{fig:BR_all}, showing the correlation between
$\tkappa$ and $\lambda$. Annihilation through a $Z$-boson occurs
in the two parallel strips with $\tkappa \approx 0.065$ and
$\tkappa \approx 0.07$. They are divided by the actual on-shell
condition $2 m_{\lsp} = m_Z$, for which LSP annihilation becomes too
efficient.

Following Eq.\eqref{eq:higgs_matrix} and replacing $\Alambda$ through
Eq.\eqref{eq:decoup_higgs} we see that the $H_1$ mass increases with
$\lambda$ directly, as well as indirectly via $\Akappa$.  Larger
values of $\lambda$ lead to a steeper increase of the scale dependence
and thereby increases $\Akappa$ at the 10~TeV scale.  At the same
time the neutralino mass increases with $\tkappa$.  This
explains why for an annihilation via $H_1$ we find a strip increasing
from $\tkappa \approx 0.02$ and $\lambda \approx 0.24$ to
$\tkappa \approx 0.06$ and $\lambda=0.4$

%----------------------------------------------------
\begin{figure}[t]
  \centering
  \includegraphics[width=0.32\textwidth]{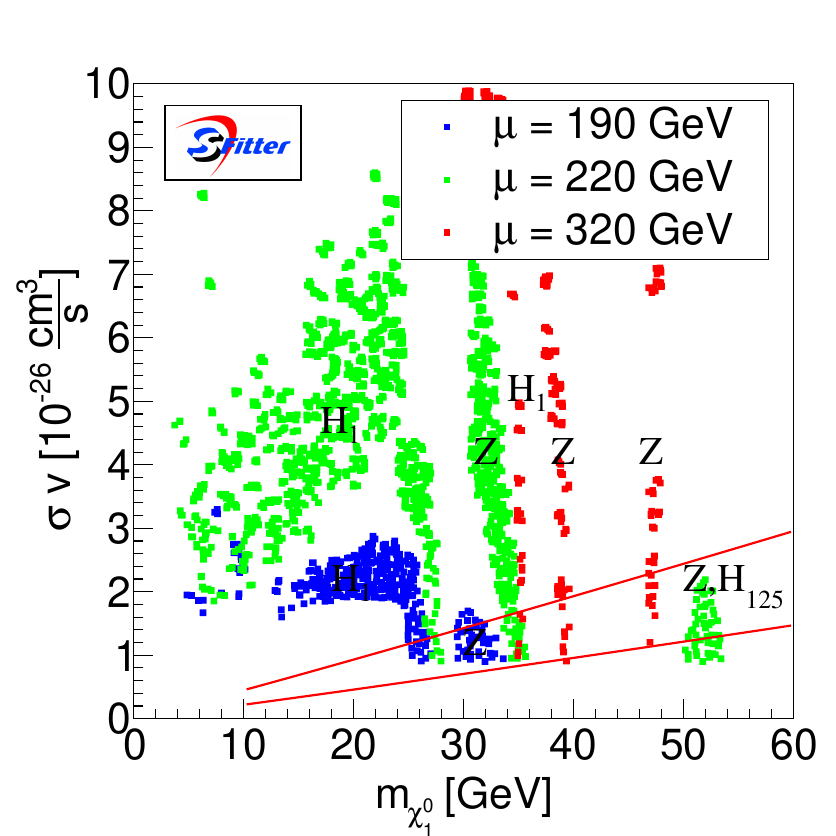}
  \caption{Correlation between the LSP mass and the dark matter
    annihilation rate for our three reference points in $\mu$, to be
    compared to for example Fig.~3 in
    Ref.~\cite{calore_cholis_weniger}. The lines correspond to the two
    sigma limits there. If available, we indicate the leading LSP
    annihilation channels. Points with $m_{\lsp} < 30$~GeV shown in
    this figure are not included in our analysis and in
    Figs.~\ref{fig:BR_190} and \ref{fig:BR_all}.}
  \label{fig:hooperon_corr} 
\end{figure}
%----------------------------------------------------

For the galactic center excess the annihilation via the pseudoscalar
$A_1$ is crucial. The LSP mass decreases 
with $\kappa$, while the mediator mass $m_{A_1}$ decreases with
$\lambda$. This is caused by the renormalization group running, which
for larger $\lambda$ pushes $\Akappa$ to larger values at
10~TeV. Following Eq.\eqref{eq:pseudo_matrix} the pseudoscalar mass
includes a factor $-\Akappa$, which means it indeed decreases with
increasing $\lambda$.  To maintain the relation between the LSP and
mediator masses, $\lambda$ and $\tkappa$ have to be
anti-correlated. This is what we observe in the two $A_1$-mediated
strips in the central lower panel of Fig.~\ref{fig:BR_all}. These
strips with the efficient pseudoscalar mediator also accommodate the
galactic center excess, as expected from the simplified model
analysis~\cite{hooperon_simplified}. Again, as before the
$A_1$-mediated annihilation blends in with $Z$-mediated annihilation.

For $\mu=320$~GeV we find an increasingly small number of parameter
points which accommodate the galactic center excess as well as the
current relic density. However, they reside in a regime with 
\begin{align}
\br(H_{125}\rightarrow \lsp \lsp) \lesssim  15\% 
\qqquad \text{for} \quad \mu = 320~\gev \; ,
\end{align}
again consistent with the galactic center excess. One aspect which
will become relevant when we embed the NMSSM in a unified framework at
the GUT scale is the $M_1$-dependence of the Hooperon-compatible
parameter points. For all three $\mu$-values we observe a tail towards
large $M_1$ values, for which the LSP properties do not change any
longer. While we do not show values above $M_1 > 90$~GeV, we could
extend the curves to much larger bino masses.\bigskip

For a combination of the different $\mu$ values assumed in the above
analysis we show the two-dimensional correlations between the LSP mass
and the annihilation rate in Fig.~\ref{fig:hooperon_corr}. The valid
NMSSM parameter points include LSP masses below 30~GeV, which we do
not consider in our analysis of the invisible branching
fractions~\cite{hooperon,calore_cholis_weniger}. We see that most of
the NMSSM points would prefer a larger annihilation rate than the
central Hooperon value, but this annihilation rate can be accommodated
by moving the different masses slightly on and off the respective
resonance conditions. Moreover, the majority of allowed points have
LSP masses between 10 and 30~GeV, in particular for $\mu=220$~GeV. On
the other hand, for all three values of $\mu$ it is possible to enter
the preferred region taken from Ref.~\cite{calore_cholis_weniger} in
the two-dimensional $\sigma v$ vs $m_{\lsp}$ plane.

%%%%%%%%%%%%%%%%%%%%%%%%%%%%%%%%%%%%%%%%%%%%%%%%%%%%%%%%%%%%%%%%%%%%%%%%%%%%%%%%
\section{High-scale NMSSM}
\label{sec:high_scale}

Instead of using the full set of TeV-scale model parameters we can
require the NMSSM to fulfill a set of unification assumptions at large
energy scales. We start with a unified squark and slepton mass $m_0$,
a unified gaugino mass $m_{1/2}$, and a unified trilinear coupling
$A_0$ at the GUT scale~\cite{msugra}.  Furthermore, we require a
$\mathbb{Z}_3$-symmetry to remove for example the $\mu$-term and
replace it with an effective $\mu$-term induced by the singlet
VEV. Because we do not require unified Higgs masses $m_{H_u}^2,
m_{H_d}^2$, and $m_S^2$ we refer to the model as the NUH-NMSSM
(non-universal Higgs masses)~\cite{nuh_nmssm}.\bigskip

From Sec.~\ref{sec:higgs_singlet} we know that to produce a light
scalar and a light pseudoscalar, $\tkappa$ has to be small and
$\Akappa$ evaluated at 10~TeV has to range around the electroweak
scale.  The running of the different model parameters from the GUT
scale to the weak scale is described by renormalization group
equations, for example for the singlet couplings
\begin{align}
16\pi^2 \frac{d \lambda}{d \log Q^2} 
&= \frac{\lambda}{2} \, \left(
  4 \lambda^2 + 3h_t^2+3h_b^2 +h_\tau^2 
+  2\kappa^2 -g_1^2-3g_2^2 \right) \notag \\
16\pi^2 \frac{d \kappa}{d \log Q^2}  
&= \frac{\kappa}{2} \, \left(6\lambda^2+6\kappa^2 \right) \notag \\
16\pi^2 \frac{d \tkappa}{d \log Q^2}
%&= 8 \pi^2 \tkappa
%\left(
%  \frac{1}{\kappa^2}   \frac{d \kappa^2}  {d \log Q^2} 
%-\frac{1}{\lambda^2} \frac{d \lambda^2}{d \log Q^2}%
%\right) \notag \\
&=\frac{\tkappa}{2} \,
\left(
4 \tkappa^2 \lambda^2 + 2 \lambda^2  
- 3h_t^2 -3h_b^2 -h_\tau^2 
+g_1^2+3g_2^2
\right)\; .
\end{align}
The Yukawa couplings are defined as $m_f = h_f v \sin
\beta/\sqrt{2}$. The couplings $\kappa$ and $\lambda$ appear squared,
so that for a choice of signs our argument will hold for their
absolute values. If we neglect the gauge couplings, $\lambda$
increases with $Q$ and runs into a Landau pole. If we also neglect the
Yukawas, which accelerate this increase, the Landau pole is
approximately given by
\begin{align}
\lambda(Q) = \lambda_0 
             \left[ 1-\frac{\lambda_0^2}{2 \pi^2} \log{\frac{Q}{Q_0}}
             \right]^{-1/2} 
\to \infty \; .
\end{align}
For our theory to be valid up to $Q= 10^{16}$~GeV the Higgs--singlet
coupling is limited to $\lambda_0 < 0.81$ at $Q_0=1$~TeV. The large and 
also increasing top Yukawa coupling further accelerates the approach of a
strongly interacting regime, requiring $\lambda_0 \lesssim 0.5~...~0.6$
for a valid theory to the GUT scale. Assuming roughly constant
$\lambda$ and also ignoring the Standard Model Yukawa and gauge couplings,
the running singlet self-coupling $\kappa$ is given by
\begin{align}
\kappa(Q) = 
\kappa_0 \; 
\left[ 1+  \kappa_0^2
           \left(1-\left(\frac{Q}{Q_0}\right)^{\frac{3 \lambda^2}{4 \pi^2}} \right)
\right]^{-1/2} \;
\left( \frac{Q}{Q_0}\right)^{\frac{3\lambda^2}{8\pi^2}} \; .
\end{align}
The maximum value of $\kappa$ for which the theory is defined up to 
$Q= 10^{16}$~GeV ranges around
\begin{align}
\kappa_\text{max}= \frac{1}{\sqrt{\left( \dfrac{Q}{Q_0}\right)^{\frac{3\lambda^2}{8\pi^2}} -1}} 
=
\begin{cases}
0.66 \qquad \lambda &= 0 \\
0.53 \qquad \lambda &= 0.6 \; .
\end{cases}
\end{align}
In this approximation we can compute a few example values by
iterating: starting with $\lambda=0.3 = \kappa_0$ at the TeV scale we
find $\kappa=0.43$ and $\lambda = 0.61$ at $10^{16}$~GeV. The singlet
mass parameter $\tkappa$ decreases from 1.0 to 0.7.

In the running of $\tkappa$ the top Yukawa coupling enters with a
negative sign. Therefore $\kappa$ increases slower with the scale than
$\lambda$ as long as $\kappa, \lambda \ll h_t$.  If we consider larger
values as weak-scale starting points, $\kappa$ increases faster than
$\lambda$.  $\lambda=0.45 = \kappa_0$ gives $\kappa=1.7$ and $\lambda
= 1.2$ at the GUT scale, so $\tkappa$ increases from 1.0 to
1.4.\bigskip

For the component fields both, the Higgs--singlet coupling and the
singlet self-coupling come with associated mass scales. They run like
\begin{align}
16\pi^2 \frac{d \Alambda}{d \log Q^2} 
&=4\lambda^2 \Alambda + 3h_t^2A_t + 3h_b^2A_b 
+ h_\tau^2A_\tau +2\kappa^2 \Akappa +g_1^2 M_1 +3 g_2^2M_2 \notag \\
16\pi^2 \frac{d \Akappa}{d \log Q^2 } 
&=6 \lambda^2 \left( \tkappa^2 \Akappa + \Alambda \right) \notag \\
16 \pi^2 \frac{d m_S^2}{d \log Q^2} 
&= 2\lambda^2 \left( m_{H_u}^2+m_{H_d}^2+m_{S}^2+\Alambda^2 
  + 3 \tkappa^2 m_S^2 + \tkappa^2 \Akappa^2 \right)
\label{eq:rge_a}
\end{align}
The increase of these mass scales towards high energy scales clearly
does not help with the appearance of a strongly interacting
Higgs--singlet sector in the NMSSM.\bigskip

Finally, we can ask how the new NMSSM Higgs--singlet parameters affect
the running of the MSSM-like parameters. An interesting parameter in
the MSSM is the stop mixing parameter, which now runs like
\begin{align}
16\pi^2 \frac{d A_t}{d \log Q^2} &= 6h_t^2A_t + h_b^2A_b +\lambda^2 \Alambda +\frac{13}{9}g_1^2M_1 +3 g_2^2M_2 + \frac{16}{3}g_3^2M_3
\end{align}
While there will be an effect of the additional singlet on the running
of the MSSM-like parameters, its impact will be numerically small. The
only exception appears when we allow with Higgs--singlet sector to
become strongly interacting at relatively low scales, in which case
for example the stop mixing parameter will also sharply
increase.

%%%%%%%%%%%%%%%%%%%%%%%%%%%%%%%%%%%%%%%%%%%%%%%%%%%%%%%%%%%%%%%%%%%%%%%%%%%%%%%%
\subsection{Global analysis}
\label{sec:global}

From the previous discussion it is clear that there are several, more
or less distinct regions of the NMSSM parameter space, which allow us
to describe the Hooperon. This is less clear when we constrain the
model parameters through unification assumptions. Most notably, the
unification of the gaugino masses at the GUT scale links the bino mass
$M_1$ to the gluino mass, which in turn is constrained by LHC
searches~\cite{gluino}.

Before we focus on the galactic center excess and invisible Higgs
decays, it makes sense to test how well the unified NUH-NMSSM can
accommodate all other available data listed in Tab.~\ref{tab:data},
including the observed relic density and the Xenon limits on
direct detection. We also include the SM-like Higgs couplings
strengths from the \textsc{SFitter}-Higgs
analysis~\cite{sfitter_higgs}. Our parameters of interest are
$\lambda$, $\tkappa$, and $\mu$.  \bigskip

In the NUH-NMSSM with decoupled scalars $(m_0 = 2~\tev)$ we face two
major differences compared to the TeV-scale model. First, unification
links the binos mass $M_1$ to the gluino mass, which is constrained by
direct LHC searches~\cite{gluino}.  In addition, the SUSY-breaking
singlet parameters $\Alambda$ and $\Akappa$ are set at the GUT
scale. This means that we cannot apply the simple decoupling relations
for example given in Eq.\eqref{eq:decoup_higgs}.

To get some control over the parameters we start with an
\textsc{SFitter} likelihood analysis, for example to see how the
requirement $H_2=H_{125}$ translates into the high-scale parameters
$\Alambda, \Akappa, m_{1/2}$, and $A_0$.  The unified gaugino mass is
proportional to the gluino mass and hence constrained to be $m_{1/2}
\gtrsim 500$~GeV. The mass scales $\Alambda$ and $\Akappa$ will
eventually be constrained by the requirement of a light scalar mass,
and we fix their ranges to $\Akappa = -1.5 ... 1.5$~TeV and $\Alambda
= -1 ... 5$~TeV.  From our experience with the TeV-scale NMSSM we
limit the singlet parameters, still defined at 1~TeV, to $\lambda <1$,
$\kappa<1$, and $\mu<400$~GeV. The ratio of the VEVs is fixed again to
$\tan \beta = 40$.

\begin{figure}[t]
  \includegraphics[width=0.48\textwidth]{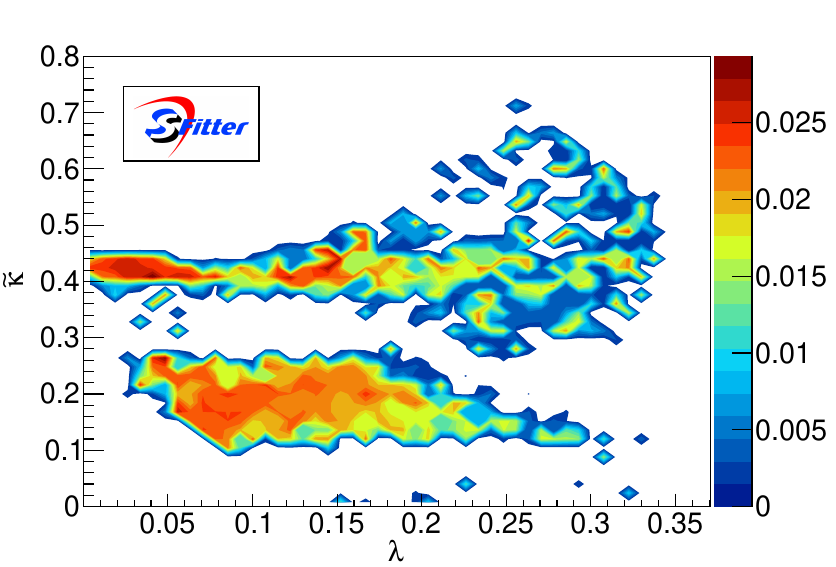}
  \includegraphics[width=0.48\textwidth]{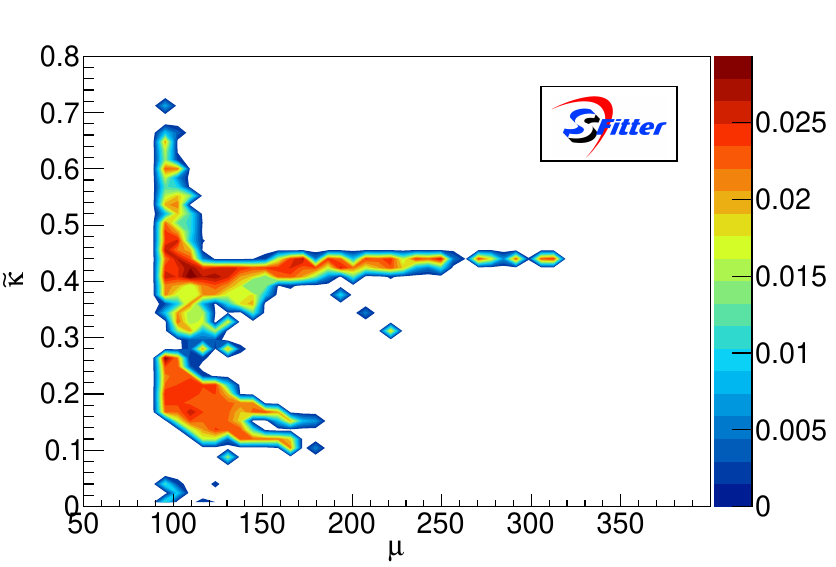}\\
  \includegraphics[width=0.48\textwidth]{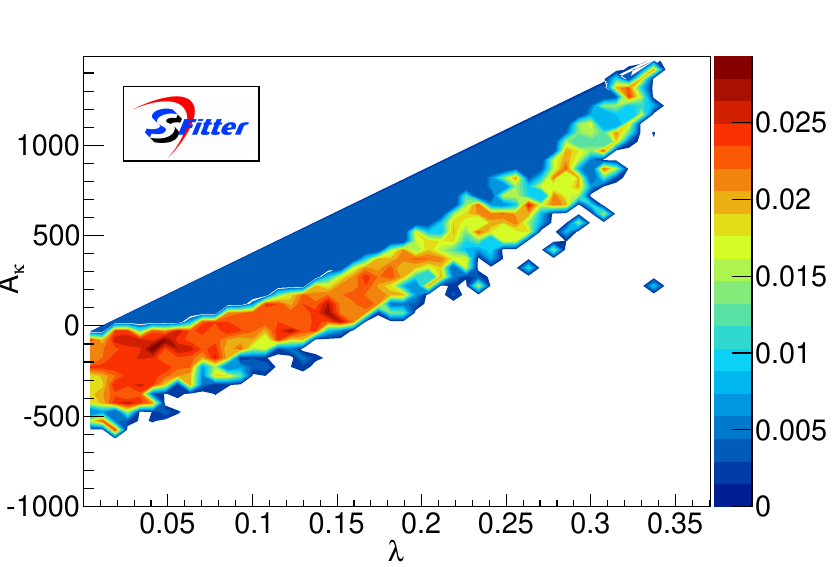}
  \includegraphics[width=0.48\textwidth]{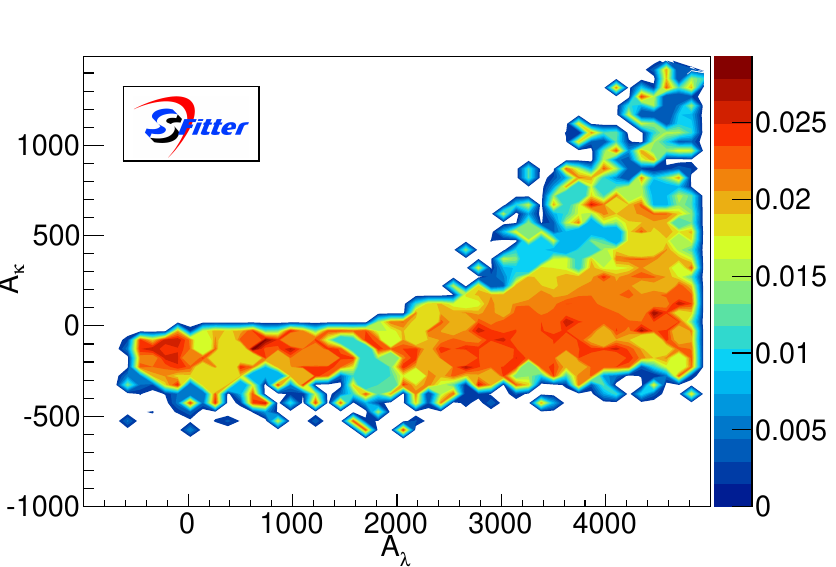}
  \caption{Profile likelihood projections of the NUH-NMSSM assuming a
    SM-like $H_2 \equiv H_{125}$. All measurements shown in
    Tab.~\ref{tab:data} and the SM-like Higgs
    couplings~\cite{sfitter_higgs} are included. Of the shown
    parameters $\Akappa$, and $\Alambda$ are defined at
    the GUT scale. The singlet parameters $\lambda, \tkappa$ and $\mu$
    are defined at 1~TeV, just like in the low scale
    scenario.}
  \label{fig:cnmssm_SMh2} 
\end{figure}

As before, we use all data listed in Tab.~\ref{tab:data}. In the Higgs
sector we identify the observed SM-like Higgs with the second-lightest
NMSSM Higgs and include the Higgs couplings measurements from ATLAS
and CMS searches~\cite{sfitter_higgs}. A set of two-dimensional
profile likelihood projections is displayed in
Fig.~\ref{fig:cnmssm_SMh2}.\bigskip

%----------------------------------------------------
%----------------------------------------------------

The upper left panel shows the profile likelihood projection on the
$\tkappa - \lambda$ plane.  We identify three distinct regions through
their dark matter annihilation channels~\cite{sfitter_planck}:
\begin{enumerate}
\item a broad band with $\tkappa= 0.1 ... 0.3, \lambda<0.25$, and
  $\mu=90 ... 200$~GeV. It includes two LSP mass regions with an
  annihilation through $Z$- and $H_{125}$-exchange.
\item a narrow strip around $\tkappa \approx 0.42, \lambda < 0.2$, and
  $\mu=90 ... 350$~GeV. It relies on a light chargino either for
  co-annihilation or for $t$-channel exchange for efficient dark
  matter annihilation.
\item a bulk region with $\tkappa = 0.3 ... 0.7, \lambda>0.2$, and
  $\mu=90 ... 150$~GeV. Here, the annihilation occurs through a mix of
  channels, notably including the light singlet pseudoscalar.
\end{enumerate} 
The transition between the second and third region is not uniquely
defined, but involves it the appearance of the $A_1$-funnel
annihilation and a drop in the LSP singlino content from
$70\%~...~90\%$ to $10\%~...~70\%$. Following Sec.~\ref{sec:hinv}, an
invisible branching ratio needs a LSP mass smaller than
62~\gev. Therefore the only region compatible with the galactic center
excess and an invisible branching ratio will be $\tkappa =
0.1~...~0.25$.\bigskip

The upper right panel in Fig.~\ref{fig:cnmssm_SMh2} shows the profile
likelihood projection on the $\tkappa - \mu$ plane. A lower bound
$\mu > 90$~GeV arises from the chargino mass limit. An upper bound
is connected to the requirement that the second-lightest NMSSM Higgs
be the SM-like state: the mass of the singlet-like Higgs is
proportional to $\mu$, so for large $\mu$ it approaches 125~GeV. This
translates into the globally observed $\mu<400$~GeV.

We then combine the range in $\mu$ with the $m_{1/2}$ dependence.  As
mentioned before, the gluino bound sets a lower limit on $m_{1/2}>500
$~GeV.  The combination of $\mu<400$~GeV and $m_{1/2}>500$~GeV results
in a sum of the bino and wino LSP components to be less than 1\%
throughout the plane. The mass and composition of the
Higgsino-singlino LSP is set by $\mu, \tkappa,$ and $\lambda$.  For
$\tkappa>0.5$ it is mainly Higgsino, with its mass set by $\mu$. For
$\tkappa<0.4$ the LSP is mainly singlino, and following
Eq.\eqref{eq:mass_approx} its mass is given by $2 \tkappa \mu$ .
Large values of $\lambda$ lead to a stronger mixing between singlino
and Higgsino.\bigskip

Of the list of regions introduced above we first consider the band
with $\tkappa = 0.1 ... 0.3$, where the singlino component is larger
than 85\%.  The corresponding values of $\mu$ range from 90~GeV to
200~GeV, resulting in two regions of neutralino mass compatible with
the measured relic density: for $m_{\lsp}= 40~...~50 $~GeV the
annihilation is mediated by a $Z$-boson, while for $m_{\lsp} =
55~...~60 $~GeV the LSP annihilates via the SM-like Higgs into $b \bar
b$ and partially into light Higgs bosons. Annihilation via the
lightest pseudoscalar, as relevant for the galactic center excess can occur, but it is
not a main annihilation channel.

In the narrow second strip, the higher value of $\tkappa=0.42$ leads
to a smaller Higgsino component, that varies between 70\% and
90\%. The higher value of $\tkappa$ leads to a mass ratio of about
0.85 between the LSP and the Higgsino-like chargino. This opens the
chargino co-annihilation channel or neutralino annihilation into $WW$
and $ZZ$ through a chargino in the $t$-channel. In the upper right
panel we can verify that this channel is open up to $\mu=350$~GeV.

Finally, for $\lambda>0.2$ the different annihilation processes are no
longer well separated. Inversely correlated to $\tkappa = 0.3~...~0.7$
the singlino component decreases from 70\% to 10\%.  For this region
we find an upper bound of $\mu< 150$~GeV, resulting in neutralino
masses between 60 and 100~GeV.  For neutralinos around 60~GeV the
annihilation proceeds via the SM-like Higgs, while for larger masses
the annihilation channel is a mixture of a pseudoscalar funnel, chargino
co-annihilation, and $t$-channel annihilation via a chargino.\bigskip

The lower panels of Fig.~\ref{fig:cnmssm_SMh2} show the profile
likelihood projection onto the $\lambda-\Akappa$ and the
$\Alambda-\Akappa$ planes. From Eq.\eqref{eq:higgs_matrix} and
Eq.\eqref{eq:pseudo_matrix} we know that the scalar and pseudoscalar
singlet mass terms increase with $\lambda$. The singlet-like scalar
has to remain lighter than 125~GeV, leading to an upper limit on
$\lambda$ depending on $\Akappa$ and $\Alambda$.  Large values of
$\Akappa$ can either push the scalar mass to too large values or lead
to a vanishing pseudoscalar mass. Both effects set an upper limit on
$\Akappa$. Because $\Akappa$ is set at the GUT scale, small starting
values of $\Akappa$ can turn negative towards the TeV scale, leading to
a very light scalar. The correlation between $\lambda$ and $\Akappa$
occurs because for large $\lambda$ we need larger values of $\Akappa$
to keep the pseudoscalar singlet heavy enough.

In the right panel we see that large values of $\Akappa$ are only
possible for even larger values of $\Alambda$.  The scalar and the
pseudoscalar singlet mass-squares differ by $\Delta m^2= 4(\Akappa
\tkappa \mu + \tkappa^2 \mu^2)$, neglecting the subleading term
proportional to $m_Z^2 s_{2 \beta} \lambda^2 \tkappa$.  For both, the
scalar and the pseudoscalar masses to be above zero, this mass
difference cannot be larger than the actual mass scale. This means that
large $\Akappa$ has to be accompanied by even larger $\Alambda$.

%%%%%%%%%%%%%%%%%%%%%%%%%%%%%%%%%%%%%%%%%%%%%%%%%%%%%%%%%%%%%%%%%%%%%%%%%%%%%%%%
\subsection{Galactic center excess}
\label{sec:cnmssm_gce}

In the TeV-scale NMSSM a singlino-like LSP with a small Higgsino
component can generate the galactic center excess in agreement with
the relic density and linked to an enhanced branching ratio $H_{125}
\rightarrow \lsp \lsp$. By definition, the NUH-NMSSM contains only a
subset of the NMSSM models: the unification condition on $m_{1/2}$
impacts the range of $M_1$, and the stop masses can no longer be set
independently of the remaining sfermion masses.

To be consistent with the TeV-scale study we decouple the sfermion
sector at $m_0=10$~TeV. This is compatible with the observed Higgs
mass, when we adjust $\Alambda$ accordingly. Again, $\tan \beta$ is
set to 40, to provide a large coupling between the pseudoscalar Higgs
and the down-type quarks.  As before, $\mu$ is set at the SUSY scale
of 1~TeV and limited to $(150~...~220)$~GeV which will be compatible
with galactic center excess. To generate a singlino mass around 40 to
50~GeV, we vary $\tkappa = 0.06~...~0.18$, following
Eq.\eqref{eq:mass_approx}.\bigskip

As mentioned in the previous section, gaugino mass unification
correlates the bino, wino, and gluino masses. Direct gluino searches
set a lower limit of $m_{1/2} > 500$~GeV~\cite{gluino}.  This leads to
a heavy wino mass, out of reach for LEP2, and defines a lower bound
$M_1 > 200$~GeV. Both, the bino and wino components of the lightest
neutralino become negligible. To compensate for the missing bino
component, the Higgsino component needs to be slightly enhanced with
respect to the TeV-scale model, leading to the slightly reduced values
of $\mu$ quoted above.

In the TeV-scale case we fix $\Alambda$ using
Eq.\eqref{eq:decoup_higgs}. In the NUH-NMSSM this is no longer
possible, as $\Alambda$ is now defined at the averaged squark mass
scale, where also the Higgs masses are computed. We can estimate that
for $\mu=200$~GeV the value of $\Alambda$ at 10~TeV has to be
approximately 8~TeV. Neglecting all contributions but $\Alambda$
itself in Eq.\eqref{eq:rge_a}, the value of $\Alambda$ increases to
around 8.6~TeV when evaluated at the GUT scale.  From the global
analysis we know that $\Akappa$ tends to have the same sign as
$\Alambda$. In this case, $\Akappa$ further increases the preferred
value of $\Alambda$ at the GUT scale to around 9~TeV. This fixed value of $\Alambda$ now 
translates into preferred ranges of $\Akappa$ and $\lambda$ via the singlet
scalar and pseudoscalar squared mass terms, which need to be larger than
zero.  Choosing $\lambda = 0.25~...~0.45$ gives 
$\Akappa = (1.5~...~5)$~TeV.\bigskip

%----------------------------------------------------
\begin{figure}[t]
  \includegraphics[width=0.32\textwidth]{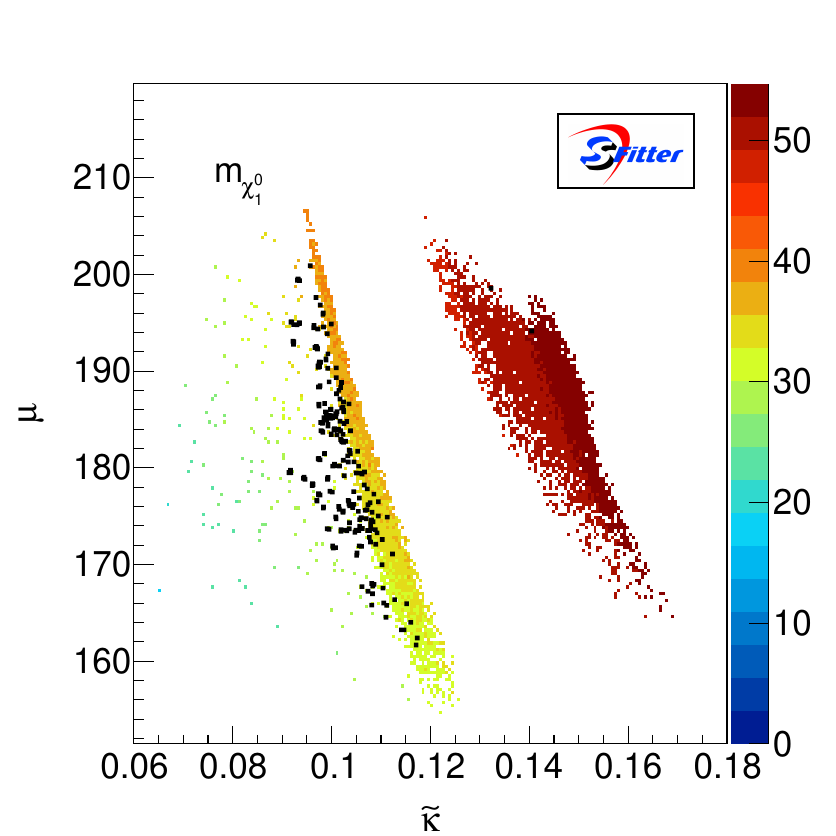}
  \includegraphics[width=0.32\textwidth]{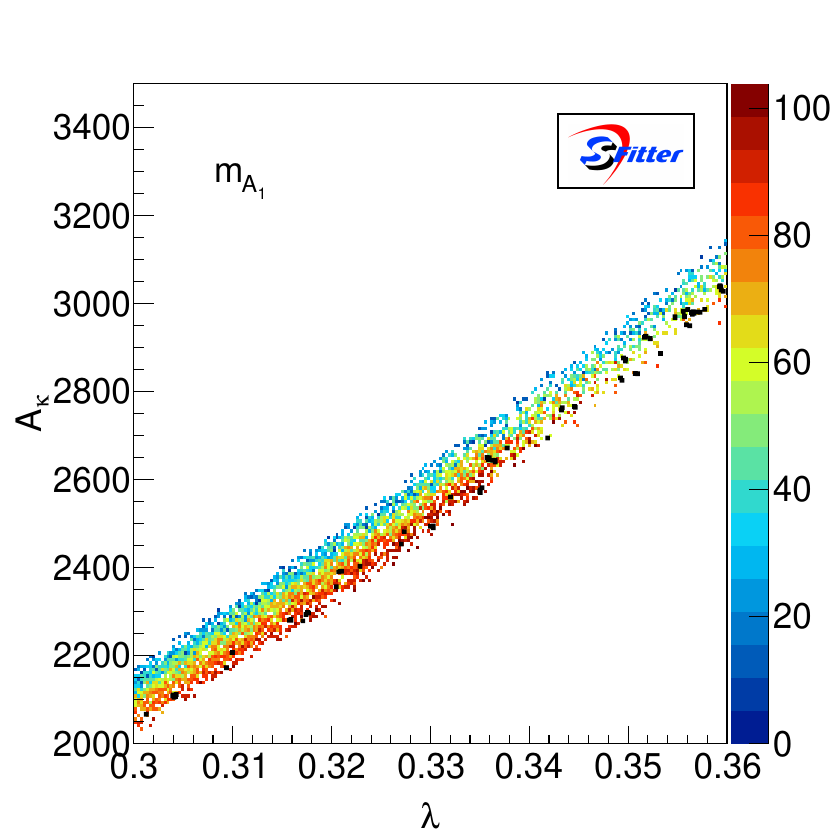}
  \includegraphics[width=0.32\textwidth]{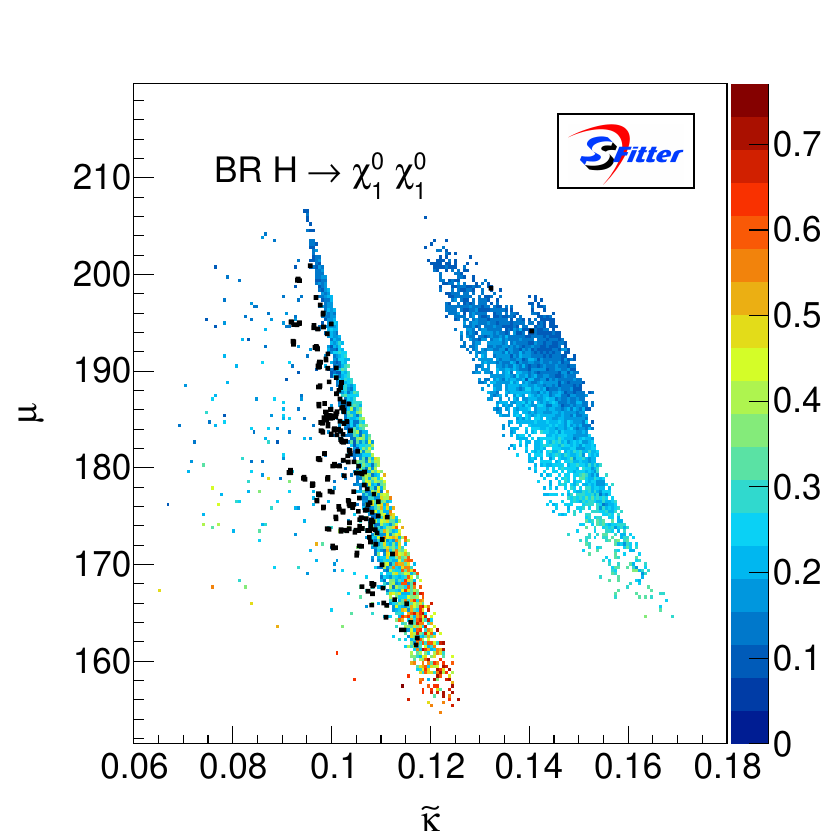}
  \caption{LSP mass, pseudoscalar mediator mass, and invisible Higgs
    branching ratio in the NUH-NMSSM.  As before, we fix $\tan
    \beta=40$. Like in Fig.~\ref{fig:BR_190} all displayed points are
    compatible with the relic density, Xenon, a chargino mass above
    103~GeV, and the correct SM-like Higgs mass. Moreover, they always
    have an invisible branching ratio $\br (H_{125} \rightarrow \lsp
    \lsp) > 10\%$. The black points are consistent with the galactic
    center excess.}
  \label{fig:cnmssm_SMh2_gce} 
\end{figure}
%----------------------------------------------------

In Fig. \ref{fig:cnmssm_SMh2_gce} we show the results of our
\textsc{SFitter} analysis.  Just as in the TeV-scale study, we require
all points to be compatible with direct detection limits, Higgs mass
measurements, and the relic density within the theoretical
uncertainty, as given in Tab.~\ref{tab:data}. The chargino masses have
to be larger than 103~GeV and the additional contribution to the
$Z$-width smaller than 2~MeV. All displayed points are consistent with
an invisible branching ratio of at least 10\%.

As mentioned before, we now study singlino LSPs with a small Higgsino
component.  On the left hand side of Fig.~\ref{fig:cnmssm_SMh2_gce} we
show the projection onto the $\tkappa -\mu$ plane. This determines the
mass and the composition of the LSP. For the singlino-like LSP the
mass increases with $\mu$ and $\tkappa$.  The allowed region for
larger $\tkappa$ corresponds to an LSP mass of 50 to 52~GeV while the
strip at $\tkappa \approx 0.11$ corresponds to a neutralino mass of 30
to 40~GeV. In between the two regions, the annihilation via the
$Z$-pole becomes too efficient.  For larger masses the annihilation is
too weak to predict the measured relic density. In contrast, towards
smaller masses a combination of the $A_1$- and $Z$-channels gives the
correct relic density as well as an annihilation cross section
compatible with the galactic center excess.

Apart from its mass, the composition of the LSP plays a key role. For
small $\tkappa \approx 0.1$ the sum of the Higgsino components
decreases with increasing $\mu$, starting from 5\% at $\mu=205$~GeV
and reaching 20\% for $\mu = 160$~GeV.  This increased active Higgsino
component implies a larger coupling to the Higgs, which leads to an
increase of the invisible branching ratio: for $\mu<160$~GeV it can
reach up to 80\%, while for $\mu > 200$~GeV it drops below the
required 10\%. However, from the large coupling to the $Z$ there
follows a negligible annihilation via the pseudoscalar mediator,
rendering this region in-compatible with the galactic center excess.
Moreover, direct detection limits become relevant for a large
$Z$-coupling and exclude points with smaller values of $\mu$.

In the center panel, we show the $\Akappa - \lambda$ plane for a
reduced range of $\lambda= 0.30~...~0.36$. This illustrates the
dependence of the lightest pseudoscalar mass on $\Akappa$ and
$\lambda$. From Eq.\eqref{eq:pseudo_matrix} now directly follows that
the $A_1$-mass increases with $\lambda$, while it decreases with
$\Akappa$. Once $\Akappa$ becomes too large, the pseudoscalar mass
squared crosses zero, limiting the allowed region.  For the galactic
center excess $\sigma v$ only reaches sufficiently high values around
the on-shell condition for the $A_1$ funnel. From the discussion of
the $\tkappa -\mu$ plane we know that the mass range for
neutralinos compatible with the galactic center excess is restricted
to $m_{\lsp} = 30~...~48$~GeV. This translated into pseudoscalar
masses of 60 to 100~GeV.

In the right panel of Fig. \ref{fig:cnmssm_SMh2_gce} the projection 
on the $\tkappa - \mu$ plane shows the resulting branching ratio for 
invisible Higgs decays. For the region around $\tkappa \approx 0.15$, 
$\lambda$ ranges from 0.25 to 0.3, while for 
$\tkappa \approx 0.11$ the allowed range for $\lambda$ increases up 
to 0.45 for $\mu=160 \gev$. Small values of $\lambda$ result in a 
small Higgsino component, leading to an invisible branching ratio of 10 
to 30\% in the region with $m_{\lsp} \approx 50 \gev$. For the narrow 
region the lower limit of $\mu =155 \gev$ in combination with large 
values of $\lambda$ allow for large invisible branching ratios up to 80\%.

When we consider only points compatible with the GCE we find

\begin{align}
\br(H_{125}\rightarrow \lsp \lsp) \lesssim 40\%
\qqquad \text{for} \quad \mu = 160 ... 200 ~\gev \; ,
\end{align}

The maximal found branching ratio of 40\% is comparable to the results 
for $\mu=220$~GeV in the TeV scale NMSSM where the bino component 
enhances the coupling. Even though the NUH-NMSSM pushes the 
neutralino content to a pure singlino-Higgsino state, we can still find 
regions that are compatible with the relic density, the GCE and a strongly 
enhanced invisible branching ratio.

%%%%%%%%%%%%%%%%%%%%%%%%%%%%%%%%%%%%%%%%%%%%%%%%%%%%%%%%%%%%%%%%%%%%%%%%%%%%%%%%
\section{Outlook}

A natural explanation of the Fermi galactic center excess is
a light, weakly interacting dark matter particle decaying to a pair of
bottom quarks through an $s$-channel pseudoscalar. The NMSSM is one of
the few models which predict precisely this process.

In the NMSSM framework, the galactic center excess as well as the
currently observed relic density can be accommodated with the help of
$Z$-funnel and $A_1$-funnel
annihilation~\cite{hooperon_scan}. Different preferred parameter
spaces can be linked to LHC searches for
trileptons~\cite{hooperon_china} or exotic Higgs
searches~\cite{papucci_zurek}. We show that for a mixed
bino--singlino--Higgsino LSP the explanation of the galactic center
excess typically predicts large invisible branching ratios of the
SM-like Higgs boson. In particular for small $\mu$ values the
invisible branching ratios can reach 50\%, testable with Run~I Higgs
data. Future LHC analyses, sensitive to invisible branching ratios
around 3\%~\cite{jamie}, cover a large fraction of Higgs decays to
a pair of Hooperons. The preferred NMSSM parameters at the TeV scale
can also be realized in a unified version of the NMSSM, albeit with
larger values of $M_1$ and slightly reduced $\mu$.\bigskip

\begin{center}
{\bf Acknowledgments}
\end{center}

AB would like to thank the Heidelberg Graduate School for Fundamental
Physics for her PhD funding, LAL for their hospitality and funding
during her master thesis work, and the DFG Graduiertenkolleg
\textsl{Particle Physics beyond the Standard Model} (GK1940). She also would 
like to thank Laurent Duflot for lively discussions on the phenomenology of the 
NMSSM and technical support at all stages of the project. TP acknowledges the 
support by the DFG Forschergruppe \textsl{New Physics at the LHC} (FOR2238).

\clearpage
%%%%%%%%%%%%%%%%%%%%%%%%%%%%%%%%%%%%%%%%%%%%%%%%%%%%%%%%%%%%%%%%%%%%%%%%%%%%%%%%
\appendix

\section{Invisible Higgs decays in the MSSM}
\label{sec:mssm}

For the sake of completion we briefly review the constraints on
invisible Higgs decays in the MSSM with a SM-like light Higgs.  The
LSP can be a combination of bino, wino, and Higgsino. First, invisible
Higgs decays $H \rightarrow \lsp \lsp$ require the LSP to be lighter
than 63 GeV.  Therefore, at least one of the mass parameters $\mu,
M_1$, and $M_2$ has to be around 100~GeV or below. The second
ingredient is the size of the coupling. Its form given in
Eq.\eqref{eq:gMSSM} requires a mixed Higgsino-gaugino states, which
means we again expect to need small values of $|\mu|$.\bigskip

%----------------------------------------------------
\begin{figure}[t]
  \includegraphics[width=0.32\textwidth]{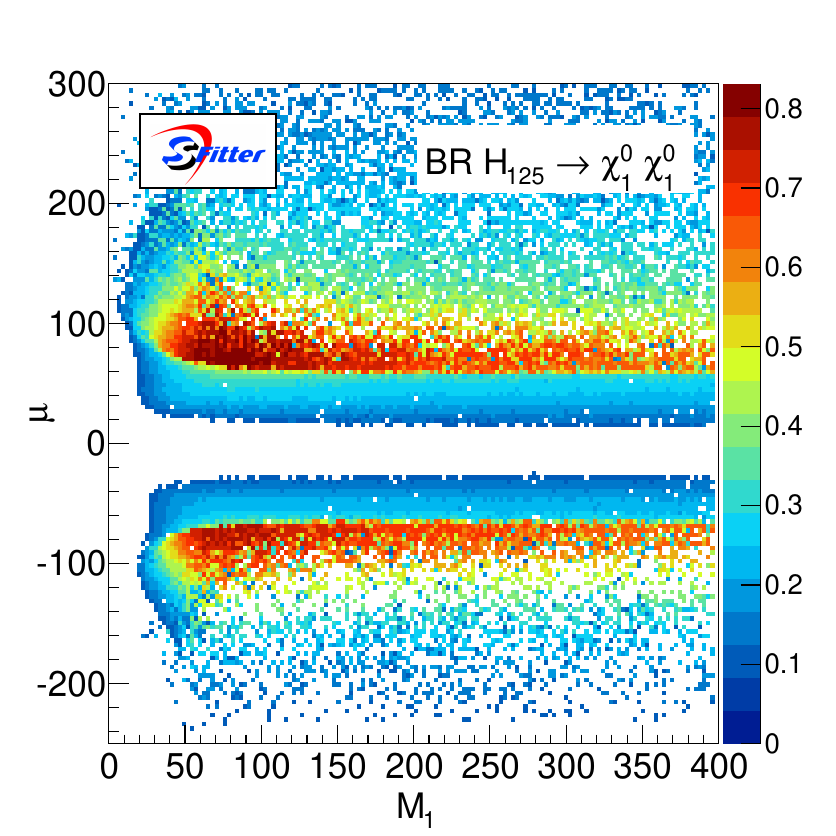}
  \includegraphics[width=0.32\textwidth]{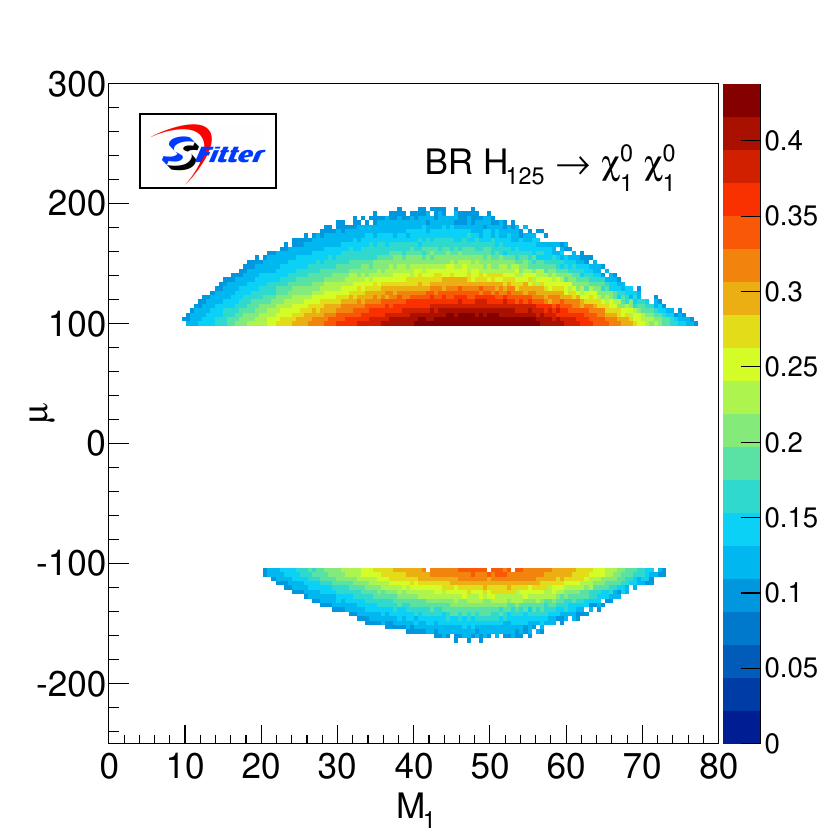}
  \includegraphics[width=0.32\textwidth]{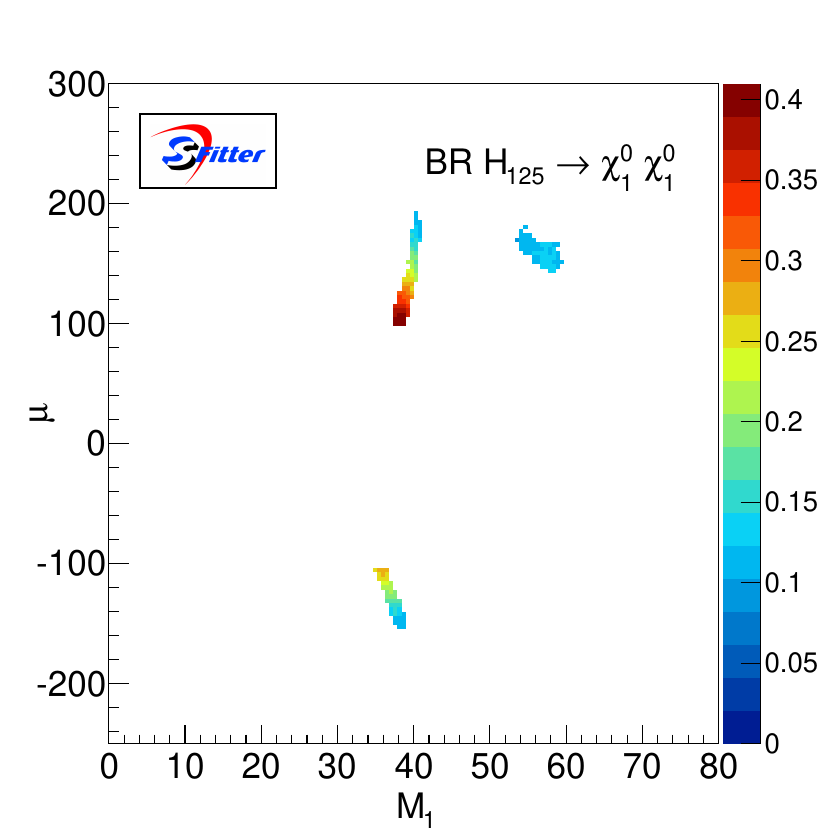}\\
 \caption{Branching ratio for $H_1\rightarrow \lsp \lsp$. From left to
   right we consecutively apply the constraints: (1)
   $m_{H_{125}}=(122~...~128)$~GeV and $\br_\text{inv}> 10\%$, (2)
   $m_{\tilde\chi^{\pm}}>103$~GeV, and (3) $\Omega_{\lsp} h^2
   =(0.107~...~0.131)$. The projections are profiled over $M_2$, showing
   the maximal branching ratio.}
  \label{fig:BRMSSM} 
\end{figure}
%----------------------------------------------------

The left panel of Fig.~\ref{fig:BRMSSM} illustrates the dependence of
the invisible branching ratio on $\mu$ and $M_1$. We vary the MSSM
parameters $\mu, M_1$, and $M_2$. For this example we set the sfermion
and gluino masses to 2 TeV, so that they decouple from the electroweak
sector, and as before set $\tan \beta = 40$.  The Higgs-sector
parameters $M_A, A_t$ and $A_\tau$ now have to be carefully chosen to
reproduce the observed Higgs mass.  For all points shown we require a
Higgs invisible branching ratio to be at least 10\%.

Without constraints on the chargino mass and dark matter properties
the maximal branching ratio exceeds 80\%. The relevant parameter space
is located around $\abs{\mu}=80$~GeV and $M_1=100~...~150$~GeV.  Away
from this region, the LSP is no longer well tempered, reducing the
coupling and leading to an invisible branching ratio below
10\%.\bigskip

In the center panel of Fig.~\ref{fig:BRMSSM} we add the LEP
limits~\cite{lep_constraints} on the chargino mass. The minimal value of
103~GeV translates into a lower bound $\mu, M_2 \gtrsim 100$~GeV. The
lower bound on $\mu$ can be seen directly in
Fig.~\ref{fig:BRMSSM}. The constraint on $M_2$ works indirectly: it
excludes wino LSPs lighter than 63~GeV, which means that a light LSP
has to be mainly bino. This is visible as an upper bound $M_1 <
80$~GeV. 

In addition to the LSP mass, we also have to adjust the couplings.
The required bino-Higgsino mixing restricts the allowed parameter
region to $M_1<80$~GeV and $\abs{\mu} <200$~GeV. Constant invisible
branching ratios correspond the two half-circles in the center panel
of Fig.~\ref{fig:BRMSSM}. Without the mass constraints on the chargino
and the Higgs we would see two approximately circular shapes centered
around $M_1 \approx 50$~GeV and slightly bigger values of $|\mu|$. This
reflects the preference for a light, well-tempered LSP with roughly
equal bino and Higgsino fractions. In particular the chargino mass
limit simply removes the region with $|\mu| \lesssim 100$~GeV.
The maximum invisible branching ratio in the MSSM is 45\%.

In the right panel of Fig.~\ref{fig:BRMSSM} we add the Planck
measurement of the LSP relic density.  In this configuration the
annihilation proceeds via an $s$-channel $Z$-boson. Planck excludes
neutralino masses around 45~GeV, where resonant annihilation leads to
a too small relic density. Three distinct strips remain compatible
with the measured relic density: two with neutralino masses between
35~GeV and 40~GeV, and one between 50~GeV and 55~GeV. However, the
latter is excluded by the Xenon100.  This additional constraint
further reduces the maximal invisible branching ratio
\begin{align}
\br (H_{125} \to \lsp \lsp) \lesssim 50\% 
\qqquad \text{for} \quad \mu = 100~\gev, \quad M_1 = 45~\gev \; ,
\end{align}
As discussed in Sec.~\ref{sec:hooperon}, in the absence of a light
pseudoscalar mediator we do not consider constraints from the galactic
center excess in the MSSM.\bigskip

Finally, we consider $Z$-decays to neutralinos for neutralino masses
smaller than 45~GeV.  The corresponding partial with adds to the width
from $Z$-decays into neutrinos, whose SM prediction already exceeds
the measured value of $ \Gamma (Z\rightarrow \text{inv})$ by 1.9~MeV.
While an invisible Higgs branching ratio of 10\% only adds an
additional 0.2~MeV to the $Z$-width, a Higgs branching ratio of 40\%
can imply an additional $Z$-decay width of 3~MeV.  This increases the
already existing tension between theory prediction and experimental
results for invisible $Z$-decays.

%%%%%%%%%%%%%%%%%%%%%%%%%%%%%%%%%%%%%%%%%%%%%%%%%%%%%%%%%%%%%%%%%%%%%%%%%%%%%%%

\end{document}